\documentclass[%
 aip,
 apl,
amsmath,amssymb,
preprint,
floatfix
]{revtex4-1}

\usepackage{graphicx}
\usepackage{dcolumn}
\usepackage{bm}
\usepackage{tikz}
\usetikzlibrary{arrows.meta}
\usepackage[utf8]{inputenc}
\usepackage[T1]{fontenc}
\usepackage{mathptmx}
\usepackage{etoolbox}
\usepackage{xcolor}
\usepackage{subcaption}
\usepackage{booktabs}
\usepackage{ragged2e}
\makeatletter
\def\@email#1#2{%
 \endgroup
 \patchcmd{\titleblock@produce}
  {\frontmatter@RRAPformat}
  {\frontmatter@RRAPformat{\produce@RRAP{*#1\href{mailto:#2}{#2}}}\frontmatter@RRAPformat}
  {}{}
}%
\makeatother
\begin{document}

%\preprint{AIP/123-QED}

\title{Automated Dislocation Detection in Electron Channelling
  Contrast Imaging: A Comparative Study of Rule-Based, Neural Network,
  and Deep Learning Approaches}
\author{A. Holmes} \email{aaron.holmes.2018@strath.ac.uk}
\affiliation{Advanced Materials Diffraction Lab, Department of
  Physics, SUPA, University of Strathclyde, Glasgow G4 0NG, United
  Kingdom} \affiliation{Quantum and Light Matter Interaction Theory
  Group (QliMIT), Department of Physics, SUPA, University of
  Strathclyde, Glasgow G4 0NG, United Kingdom}
\author{C. Trager-Cowan} \affiliation{Advanced Materials Diffraction
  Lab, Department of Physics, SUPA, University of Strathclyde, Glasgow
  G4 0NG, United Kingdom} \author{J. Bruckbauer} \affiliation{Advanced
  Materials Diffraction Lab, Department of Physics, SUPA, University
  of Strathclyde, Glasgow G4 0NG, United Kingdom}
\author{B. Hourahine} \affiliation{Advanced Materials Diffraction Lab,
  Department of Physics, SUPA, University of Strathclyde, Glasgow G4
  0NG, United Kingdom} \affiliation{Quantum and Light Matter
  Interaction Theory Group (QliMIT), Department of Physics, SUPA,
  University of Strathclyde, Glasgow G4 0NG, United Kingdom}

%\date{\today}

\begin{abstract}
Quantifying threading dislocations in semiconductor materials via
electron channelling contrast imaging (ECCI) is heavily bottlenecked
by slow manual analysis. This work benchmarks three automated
detection pipelines on ECCI micrographs of gallium nitride (GaN)
against a statistical ground truth. A classical rule-based computer
vision approach proved unreliable due to extensive per-image tuning
requirements, while a convolutional neural network (CNN)-based
multi-stage classification and locator method achieved 87\% accuracy
but required significant time for a test image and was less efficient
in high-density regions. By contrast, a unified single-stage you only
look once (YOLOv8) architecture achieved a counting accuracy of
98.6\%, alongside 98.7\% precision and 98.7\% recall across 7451
dislocations over multiple images, with rapid inference
times. Successful deployment of YOLOv8 required addressing two
domain-specific machine learning challenges. First, to mitigate the
model's scale sensitivity, an adaptive gaussian tile-sizing algorithm
was developed to optimise the field of view per image. Second, the
deployed confidence threshold (0.025) differed substantially from the
suggested inference default value for YOLOv8 (0.25), raising counting
accuracy from 90.8\% to 99.3\% on a subsection of the benchmark
image. In this case, low confidence scores represented physical signal
strength rather than classification ambiguity. This unified,
scale-adaptive approach demonstrates practical viability for
high-throughput, quantitative semiconductor defect characterisation.
\end{abstract}

\maketitle
%%%%%%%%%%%%%%%%%%%%%%%%%%%%%%%%%%%%%%%%%%%%%%%%%%%%%%%%%%%%%%%%%
\section{Introduction}
%%%%%%%%%%%%%%%%%%%%%%%%%%%%%%%%%%%%%%%%%%%%%%%%%%%%%%%%%%%%%%%%%

Gallium nitride (GaN) is the foundational semiconductor for modern
blue and white light-emitting diodes (LEDs)~\cite{nakamura2015nobel}
and is increasingly replacing silicon in high-power electronics
applications, including fast chargers, electric-vehicle drivetrains,
and 5G base station power
amplifiers.~\cite{Rafin2023,meneghini2017power} This transition is
driven by the material's capacity to sustain higher breakdown
voltages, operate at higher temperatures, and achieve higher switching
frequencies than silicon.~\cite{Amano_2018,mishra2008gan} Advancing
these technologies relies upon improving the structural properties of
the underlying material. While GaN devices can tolerate varying
degrees of crystalline imperfection, their overall efficiency and
reliability are linked to the density of threading
dislocations.~\cite{Bennett01092010} These line
defects~\cite{hull2011introduction} trap charge carriers and dissipate
their energy, which reduces the overall efficiency of
LEDs,~\cite{Bennett01092010} impedes electrical flow in transistors,
and increases leakage currents in power
devices.~\cite{usami2018correlation}

\begin{figure}[tb]
\centering
\begin{tikzpicture}
    \node[anchor=south west, inner sep=0] (image) at (0,0) {
        \includegraphics[width=0.48\textwidth]{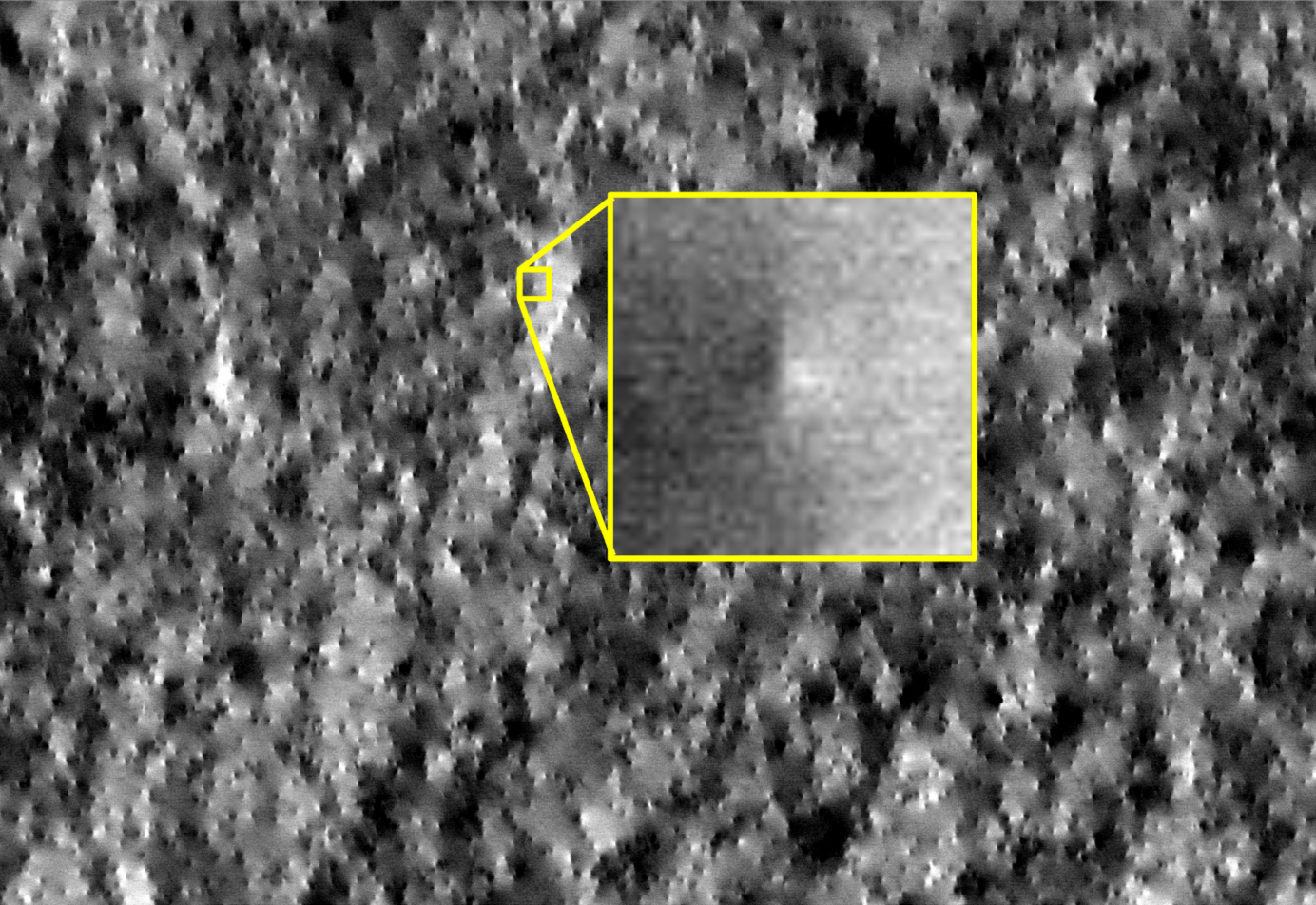}
    };
    \begin{scope}[x={(image.south east)},y={(image.north west)}]
        \draw[white, line width=1.5pt, |-|] (0.005, 0.05) -- (0.25, 0.05) 
            node[midway, above, text=white, font=\sffamily\small] {$10\,\mu\mathrm{m}$};
    \end{scope}
\end{tikzpicture}
\caption{\label{fig:ecci_example_with_zoom_int_single_dislocation} A
  representative ECCI micrograph showing dislocations as
  characteristic dipole contrast patterns. Inset: magnified view of a
  single dislocation showing the black-white contrast of the
  dipole. The varying background grey levels correspond to different
  sub-grain orientations.}
\end{figure}

Quantifying the density of threading dislocations is therefore a
fundamental step in materials characterisation. Electron channelling
contrast imaging (ECCI), in a scanning electron microscope (SEM), has
emerged as a rapid technique for this
purpose.~\cite{SIMKIN199965,hiller2026imaging} ECCI can be performed
directly on bulk samples without the need for destructive sample
preparation, resolving individual dislocations threading to the
surface as distinctive spots with black-white contrast variation
(typically as dipoles) across fields of view spanning tens of
micrometres
(Fig.~\ref{fig:ecci_example_with_zoom_int_single_dislocation}). While
data acquisition via ECCI is efficient, the subsequent analysis
remains a slow, manually intensive bottleneck. While human counting
may be manageable for isolated images containing only a few hundred
dislocations, the approach becomes infeasible for high-throughput
characterisation, where datasets contain hundreds of images with
thousands of defects. Automating this detection is therefore critical,
but it requires distinguishing subtle contrast variations from
background topography across varying dislocation densities.

Object detection is a classical problem in computer vision that has
evolved significantly. Early approaches relied on techniques such as
thresholding and edge detection,~\cite{szeliski2010computer} which
depend heavily on hand-crafted rules and parameter tuning. As
computational capabilities advanced, the field shifted toward
multi-stage convolutional neural networks (CNNs)~\cite{lecun2015deep}
that automated feature extraction but remained computationally
expensive due to their segmented classification and localisation
pipelines. More recently, unified single-stage detectors like YOLO
(you only look once)~\cite{redmon2016yolo} have revolutionised the
field by performing both tasks in a single, highly efficient forward
pass. Recognising these advantages, the materials science community is
actively exploring these modern architectures. Convolutional
approaches have been used to detect and quantify irradiation-induced
defect clusters in transmission electron microscopy (TEM) images of
steels,~\cite{li2018automated,shen2021multidefect} and to segment
dislocation lines, precipitates, and voids in scanning TEM micrographs
at accuracies comparable to human
experts.~\cite{roberts2019defectsegnet} For instance, recent work
benchmarking YOLOv8 against semantic segmentation models (like U-Net)
for transmission electron micrographs of polycrystalline films
demonstrated that YOLO architectures offer drastically increased
inference speedups suitable for real-time
analysis.~\cite{patrick2025comparative}

However, that same study identified a critical limitation when
applying the YOLOv8 architecture to microscopy. Patrick \textit{et
  al.}~\cite{patrick2025comparative} observed that YOLOv8
systematically overestimates the size of small features and
underestimates large ones when their apparent pixel diameter deviated
from a model-specific optimum (near 70 px at a $640\times640$
input). Retraining with scale-aware augmentations reduced the bias at
the extremes by approximately one third, but did not eliminate
it. This bias is likely structural: single-stage detectors (networks
that decide both what an object is and where it lies in one pass over
the image) localise objects through a small number of detection heads,
the output layers that each divide the image into a grid of cells and
predict objects within them, each operating at a fixed grid stride
with a correspondingly fixed effective receptive field (the area of
the original image feeding into a single grid cell), so features whose
apparent pixel size falls outside the scales these heads specialise in
are handled by a head with a mismatched receptive field. Crucially,
the relevant quantity is the apparent size of the feature in
pixels---set jointly by the magnification and the field of view
presented to the network---rather than the physical spatial resolution
of the measurement. Successful deployment to microscopy may therefore
require explicit control over the apparent scale of features at the
inference stage.

While this scale limitation specifically motivates careful YOLO
deployment, the broader question of which type of method is best
suited to ECCI dislocation detection remains open. This work develops
and benchmarks three detection pipelines spanning this computational
evolution: (1) a classical rule-based approach using adaptive
thresholding, requiring no training data and offering fully
interpretable decisions; (2) a custom multi-stage CNN that separates
classification and localisation, enabling targeted treatment of
distinct dislocation sub-types; and (3) a unified YOLOv8 architecture,
providing end-to-end optimisation and rapid single-pass
inference. Against a rigorously verified, human-counted ground truth,
we show not only which approach yields the highest accuracy, but also
how two domain-specific adaptations---adaptive gaussian tile-sizing to
address YOLOv8's scale bias, and recalibration of the confidence
threshold to reflect ECCI signal characteristics---are essential to
deploy single-stage detectors successfully for high-throughput defect
characterisation.

%%%%%%%%%%%%%%%%%%%%%%%%%%%%%%%%%%%%%%%%%%%%%%%%%%%%%%%%%%%%%%%%%
\section{Background and Image Characteristics}
%%%%%%%%%%%%%%%%%%%%%%%%%%%%%%%%%%%%%%%%%%%%%%%%%%%%%%%%%%%%%%%%%

\subsection{Threading Dislocations in GaN}
Gallium nitride devices are typically grown on foreign substrates
(such as sapphire or silicon)~\cite{AKASAKI1989209} due to the high
cost of native GaN wafers.~\cite{Paskova2010,fujito2009bulk}
Heteroepitaxial growth of GaN thin films leads to the formation of
threading dislocations which propagate through the crystal, ultimately
emerging at the sample
surface.~\cite{mathis2001modeling,speck1999mechanisms,moram2009origin}
In devices and LEDs, these defects act as leakage paths and
non-radiative recombination
centres,~\cite{Bennett01092010,usami2018correlation} making their
quantification a critical metric of material quality.

\subsection{Image Formation and the Dipole Signature}
ECCI allows these defects to be observed directly at/close to the
surface of bulk samples and thin
films.~\cite{SIMKIN199965,hiller2026imaging} In ECCI, the intensity of
backscattered electrons (BSE) is highly sensitive to the precise angle
between the incident electron beam and the crystal
lattice.~\cite{wilkinson1997electron,hiller2026imaging,joy1982electron}
When a region within the sample is perfectly aligned to a specific
diffraction condition, the background BSE yield is relatively uniform.

However, the misorientation and strain associated with a threading
dislocation disrupt this local alignment. The defect bends and deforms
the surrounding crystal. The misorientation and strain gradients
translate into the characteristic light--dark dipole pattern
(Fig.~\ref{fig:ecci_example_with_zoom_int_single_dislocation}) that
serves as the primary visual target for automated object
detection. Although less common in our dataset, different diffraction
conditions or regional variations can also produce a quadrupole
pattern.

\subsection{Variable Imaging Conditions and Detection Challenges}
While isolated dipoles on a uniform background should be easily
identifiable, real-world ECCI data presents a challenging computer
vision environment. Because ECCI contrast is highly sensitive to
crystal orientation and strain, even minor changes in surface
topography or orientation cause background intensity
differences.~\cite{hiller2026imaging}

This sensitivity complicates automated detection because the
visibility of any given dislocation depends on the local background
conditions, the diffraction conditions, the inherent contrast of the
defect itself, and the defect type. As the local crystal orientation
deviates from the optimal diffraction angle---approaching the edge of
the imaging conditions---the variation in BSE yield for different
sub-grains can exceed the dynamic range of the BSE detector. While the
microscope operator tunes the diffraction conditions to capture the
best overall contrast, any single field of view is inherently a
compromise. Optimising the contrast for one region often pushes
another into the noise floor or into saturation, resulting in overly
dark or bright sub-grains where the characteristic dipole signature
fades. Furthermore, some dislocations inherently produce much weaker
physical signals even in well-aligned regions. Consequently, within a
single field of view, some defects appear as high-contrast dipoles,
while others become faint, ambiguous pixel fluctuations that challenge
even human experts. Finally, dislocations frequently group together,
creating dense clusters that merge into complex, overlapping contrast
patterns.

Consequently, an automated detection system must not only navigate
scale variance but also distinguish genuine, faint physical signals
from topographic noise and be capable of dealing with high density
areas where dislocations may overlap.

%%%%%%%%%%%%%%%%%%%%%%%%%%%%%%%%%%%%%%%%%%%%%%%%%%%%%%%%%%%%%%%%%
\section{Dataset and Ground Truth}
%%%%%%%%%%%%%%%%%%%%%%%%%%%%%%%%%%%%%%%%%%%%%%%%%%%%%%%%%%%%%%%%%

The datasets used in this work consist of ECCI micrographs from a
range of GaN thin films, spanning a range of dislocation densities,
contrast conditions, sub-grain sizes, and magnification scales typical
of standard growth processes. All GaN films are grown on
\textit{c}-plane sapphire using metal-organic chemical vapour
deposition (MOCVD). The images were separated into sets for training,
validation, and testing, with a subsection of the benchmark test image
used for confidence-threshold calibration
(Table~\ref{tab:threshold_calibration}); this calibration is carried
out once and is not required for each image. Training and validation
used cropped sub-regions of these micrographs, whereas testing was
carried out on complete, uncropped images. The final unseen test set
comprises multiple full-scale images and contains a manually counted
total of 7451 dislocations. From this test set, one representative
test image
(Fig.~\ref{fig:ecci_example_with_zoom_int_single_dislocation},
$2048\times1428\,\mathrm{px}^{2}$, $3805 \pm 46$ dislocations, or a
dislocation density of $\approx3\times10^{8}$\,cm$^{-2}$) was selected
as the primary benchmark for repeated-count statistical analysis.

\subsection{Statistical Ground Truth}
\label{sec:ground_truth}
Manual counting is used as a reference for dislocation count and
location, but suffers from variability. To quantify this and establish
a statistically meaningful ground truth, the test image was counted
three times by a single operator: one full unassisted count (3790) and
two exhaustive model verification passes (3800 and 3826;
Secs.~\ref{sec:results_m2} and \ref{sec:results_m3}). These give a
sample mean $\bar{x} = 3805$ (an overbar denotes the mean throughout)
and standard deviation $s = 18.6$, with a 95\% confidence interval of
$3805 \pm 46$ dislocations using Student's
$t$-distribution~\cite{student1908} ($n=3$), and a coefficient of
variation of $0.49\%$. The corresponding coefficient of variation sets
the target accuracy that automated methods aim to approach. Additional
images received single counts, for which the relevant spread is the
prediction interval for a single observation, $\bar{x} \pm
t_{0.975,n-1}\,s\sqrt{1+1/n} = \pm 2.43\%$, rather than the confidence
interval on the mean. A conservative $\pm 2.5\%$ was therefore
adopted. Two ground truths follow from these counts. Counting accuracy
is assessed against the sample mean, $3805 \pm 46$, as the best
estimate of the true number of dislocations. Precision, recall, and F1
are instead assessed against the individual located count from the
corresponding verification pass (3800 for Method~2, 3826 for
Method~3), since matching detections to dislocations requires a
specific set of identified positions rather than an average. Aggregate
multi-image totals are likewise built from located counts. Throughout,
collinear tri-pole features (Sec.~\ref{sec:training_data}) are counted
as single dislocations, whereas kinked multi-pole contrasts, in which
the poles are not collinear
(Fig.~\ref{fig:custom_classification_dataset_examples}(f)), are
counted as two.

\subsection{Training Data}
\label{sec:training_data}
\begin{figure}[h!]
    \centering
    % --- First row ---
    \begin{subfigure}[t]{0.2\linewidth}
        \centering
        \includegraphics[width=\linewidth]{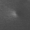}
        \caption{\emph{Single} dislocation}
    \end{subfigure}\hfill
    \begin{subfigure}[t]{0.2\linewidth}
        \centering
        \includegraphics[width=\linewidth]{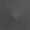}
        \caption{\emph{Faint} dislocation}
    \end{subfigure}\hfill
    \begin{subfigure}[t]{0.2\linewidth}
        \centering
        \includegraphics[width=\linewidth]{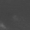}
        \caption{\emph{Ultra-faint} dislocation}
    \end{subfigure}
 
    % --- Second row ---
    \begin{subfigure}[t]{0.2\linewidth}
        \centering
        \includegraphics[width=\linewidth]{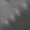}
        \caption{\emph{Multiple} dislocations}
    \end{subfigure}\hfill
    \begin{subfigure}[t]{0.2\linewidth}
        \centering
        \includegraphics[width=\linewidth]{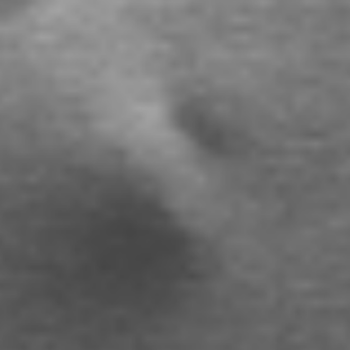}
        \caption{\emph{Tri-pole} dislocation}
    \end{subfigure}\hfill
    \begin{subfigure}[t]{0.2\linewidth}
        \centering
        \includegraphics[width=\linewidth]{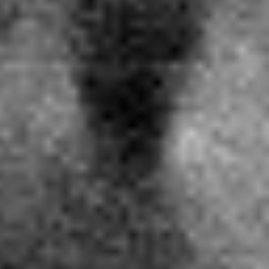}
        \caption{Kinked multi-pole}
        \label{fig:class_kinked}
    \end{subfigure}
 
    % --- Third row ---
    \begin{subfigure}[t]{0.2\linewidth}
        \centering
        \includegraphics[width=\linewidth]{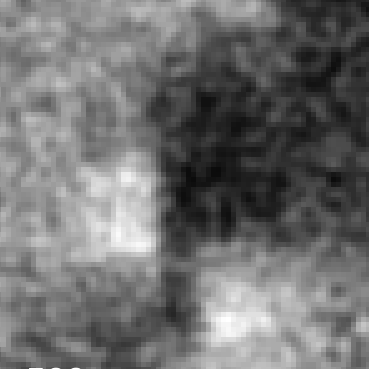}
        \caption{\emph{Quad-pole} dislocation}
    \end{subfigure}\hfill
    \begin{subfigure}[t]{0.2\linewidth}
        \centering
        \includegraphics[width=\linewidth]{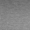}
        \caption{\emph{No dislocation}}
    \end{subfigure}\hfill
    \begin{subfigure}[t]{0.2\linewidth}
        \centering
        \begin{tikzpicture}
            \node[anchor=south west, inner sep=0] (image) at (0,0) {
                \includegraphics[width=\linewidth]{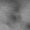}
            };
            \begin{scope}[x={(image.south east)},y={(image.north west)}]
                \draw[white, line width=1.5pt, |-|] (0.05, 0.08) -- (0.95, 0.08) 
                    node[midway, above, text=white, font=\sffamily\small] {$500\,\mathrm{nm}$};
            \end{scope}
        \end{tikzpicture}
        \caption{\emph{Unknown}}
    \end{subfigure}
 
    \caption{\label{fig:custom_classification_dataset_examples}
      Example images showing the visual differences between training
      classes. The classes are defined as follows: (a) \emph{Single}:
      clear, isolated dislocations within the central quarter; (b)
      \emph{Faint}: moderately clear, lower-contrast defects; (c)
      \emph{Ultra-faint}: barely visible contrast; (d)
      \emph{Multiple}: two or more complete dislocations in the patch,
      closely spaced or slightly overlapping, with at least one
      centred; dislocations at the patch edge are excluded unless they
      lie within or touch the central quarter; (e) \emph{Tri-pole}:
      three visibly collinear poles of similar contrast, treated as a
      single dislocation; (f) a kink in the line of poles instead
      indicates two overlapping dislocations rather than one, so such
      patches are counted as two dislocations and labelled
      \emph{multiple} rather than \emph{tri-pole}; (g)
      \emph{Quad-pole}: four poles in a cross pattern; (h) \emph{No
      dislocation}: empty windows or defects only near the edge of the
      patch; and (i) \emph{Unknown}: ambiguous contrasts, unusual
      artefacts, or highly obscure clusters. All patches represent an
      approximate 500 nm field of view.}
\end{figure}

The two learning-based methods used different training strategies. The
multi-stage CNN required $\approx 13{,}000$ small patches ($\approx
60\times60\,\mathrm{px}^{2}$, small variations in size were used to
help with generalisation) labelled across eight categories
(\emph{single}, \emph{faint}, \emph{ultra-faint}, \emph{multiple},
\emph{tri-pole}, \emph{quad-pole}, \emph{no dislocation}, and
\emph{unknown}) shown in
Fig.~\ref{fig:custom_classification_dataset_examples}, since
downstream processing was routed by class. A subset of $\approx 1000$
manually-labelled patches provided bounding-box ground truth for the
locator network.

For YOLO, $\approx 8700$ cropped sub-regions were extracted from ECCI
micrographs outside the test set, with all sub-types collapsed into a
single unified \emph{dislocation} class to exploit the architecture's
tolerance to intra-class variation. Crop sizes were varied both across
and within source images, alongside deliberate variation in density,
clustering pattern, and local contrast, exposing the model to a range
of apparent dislocation scales and spatial arrangements without
requiring explicit multi-scale training. Where possible bounding-box
labels from the outputs of the multi-stage CNN pipeline were used
after manual correction: false positives were removed, missed
detections added, and boundaries refined. The labelled set was split
90\% training and 10\% into an internal validation split, with
standard YOLOv8 augmentation applied during training (horizontal
flips, hue, saturation, and value jitter, and random translation and
scaling).

%%%%%%%%%%%%%%%%%%%%%%%%%%%%%%%%%%%%%%%%%%%%%%%%%%%%%%%%%%%%%%%%%
\section{Detection methods}
%%%%%%%%%%%%%%%%%%%%%%%%%%%%%%%%%%%%%%%%%%%%%%%%%%%%%%%%%%%%%%%%%

Three automated detection pipelines were developed and benchmarked: a
rule-based baseline (method 1), a multi-stage CNN approach (method 2),
and a unified single-stage detector (method 3). All methods were
evaluated on identical hardware (AMD Ryzen Threadripper 3960X, 128\,GB
RAM, NVIDIA RTX 3080, 10\,GB VRAM, 8704 CUDA cores).

\subsection{Method 1: Rule-Based Computer Vision}
The first method comprised three main operations in sequence:
intensity-based segmentation, sliding-window contrast detection, and
spatial clustering.

\textit{Segmentation.} A lower-biased multi-level Otsu
algorithm~\cite{otsu1979threshold} partitioned each image into $k$
intensity classes (typically 4--6). Standard two-class Otsu was first
applied; the lower-intensity class was then recursively subdivided
with $k{-}1$ additional thresholds. This concentrated segments in the
darker regions, where channelling contrast is most sensitive, whilst
grouping bright grains into a single broad class which had a wider
tolerance for variance in the next step.

\textit{Local contrast detection.} Within each segmented region, a
square sliding window traversed the image at a stride equal to half
the window size. The exact window dimension (typically $\approx
5\times 5\,\mathrm{px}^{2}$) was tuned for each image resolution, the
size was initially set large and iteratively reduced until visual
inspection confirmed most dislocations would be picked up. Local
maximum contrast is computed at each position as the pixel intensity
range ($\max-\min$). Window centres exceeding a region-specific
contrast threshold (10--100 on a standard 0--255 intensity scale
[internally normalised between 0 and 1], lower for darker regions)
were retained as candidate detections.

\textit{Clustering.} Multiple hits on a single dislocation were merged
using mean-shift
clustering,~\cite{fukunaga1975estimation,comaniciu2002mean} with the
bandwidth manually tuned per image to merge overlapping detections
whilst attempting to preserve separation between adjacent
features. Cluster centres then defined the final detection
locations. The pipeline required 5--8 manually-set parameters per
image, with 15--30 minutes of iterative refinement and processing
times of minutes to hours per image depending on size of image, size
of detection window (smaller windows create more points to cluster,
but also clearer groups), number of clusters, and density of clusters.

\subsection{Method 2: Multi-Stage CNN Pipeline}

The second method comprised three independently-trained CNNs operating
on small greyscale patches: a binary classifier, a bounding-box
regression network, and a multi-class classifier with an OpenMax
activation function.~\cite{bendale2016openmax}

\textit{Network roles.} The binary classifier (trained on all $\approx
13{,}000$ patches with simple dislocation/no-dislocation labels) acted
as an initial filter, and was found to yield more reliable detections
than using the multi-class network alone. The locator network, trained
with mean squared error (MSE) on $\approx 1000$ manually-labelled
patches, regressed to give four bounding-box coordinates $(x_{\min},
y_{\min}, x_{\max}, y_{\max})$. The multi-class network assigned one
of the eight classes, with OpenMax applied to the penultimate layer to
flag out-of-distribution inputs.

\textit{Detection pipeline.} A manually-sized scan-box (typically
$\approx 36$\,px) rasterised the image at a stride equal to one sixth
of the box size. Each patch was first screened by the binary
classifier: positive detections were passed to the locator network,
which predicted a bounding box. An iterative re-centring loop then
adjusted the scan-box position until the feature was centred within a
few pixels, up to a convergence limit of 3 local spatial iterations or
a maximum of 5 classification state changes, after which unresolved
features were flagged as \textit{unknown}. The multi-class network
then assigned a label, triggering class-specific downstream
processing: \textit{single}/\textit{faint}/\textit{ultra-faint}
patches initiated a perimeter search for nearby dislocations;
\textit{multiple} patches performed iterative neighbour searches;
\textit{tri-pole}/\textit{quad-pole} patches used gradient-based peak
detection. Because spatial gradient operations inherently amplify
high-frequency noise in the BSE/ECCI image, the image was first
pre-processed with a gaussian filter to suppress false peaks. Within
the flagged patches, local gradient maxima were identified to seed
local bounding boxes. A spatial clustering algorithm then grouped any
adjacent boxes falling within a set distance threshold, merging the
largest connected cluster into a single overarching bounding box to
ensure the complete multi-pole feature was captured. Finally,
detections falling within a fixed pixel radius of an already
established centre were rejected as duplicates, and patches that
failed to resolve were flagged as \textit{unknown} for manual review.

\subsection{Method 3: YOLOv8 with Adaptive Tile Sizing}
\label{sec:methods_yolo}

The third method used YOLOv8n,~\cite{yolov8_ultralytics} a
single-stage detector that performs classification and localisation in
a single forward pass. To mitigate the scale bias also identified by
Patrick \textit{et al.}~\cite{patrick2025comparative}, we introduce an
adaptive gaussian tile-sizing procedure that automatically rescales
each test image so that dislocations appear at the model's preferred
apparent size.

\textit{Architecture.} YOLOv8n consists of a CSPDarknet53
backbone~\cite{wang2020cspnet} for hierarchical feature extraction, a
Path Aggregation Network (PANet) neck~\cite{liu2018path} for
bidirectional multi-scale feature fusion, and three anchor-free
detection heads operating at grid resolutions of $128\times128$,
$64\times64$, and $32\times32$ at an input size of $1024\times1024$
(specialised for small, medium, and large objects respectively).

\textit{Training.} Training used the Ultralytics YOLO framework with
$1024\times1024\,\mathrm{px}^{2}$ input images, batch size 8, and
stochastic gradient descent with momentum.~\cite{yolov8_ultralytics}
The composite loss combined Complete-Intersection over Union (CIoU)
loss for bounding-box regression, binary cross-entropy for
classification, and distribution focal loss for refined boundary
prediction. While standard intersection over union
(IoU)~\cite{jaccard1912} only measures the overlap area, CIoU also
penalises offsets between box centres and differences in aspect ratio,
enforcing tighter localisation around the defect contrast
patterns. Built-in augmentation included horizontal flips,
hue/saturation/value jitter, and random translation and
scaling. Training terminated after 628 epochs via early stopping
(patience 100); the best validation checkpoint (epoch 528) was used
for all subsequent inference.

\textit{Adaptive gaussian tile sizing.} Because apparent dislocation
size varies substantially with magnification across samples, a fixed
inference tile size cannot place every dislocation within the model's
preferred detection scale. A two-stage calibration was therefore
applied to each test image.

In the \textit{coarse} stage, up to eight
$200\times200\,\mathrm{px}^{2}$ windows were sampled across the image
and processed with the trained model. The mean detected dislocation
size, $\bar{d}$, from these windows gave an initial tile size chosen
to be $t_{\mathrm{coarse}} = 10\bar{d}$.

In the \textit{refinement} stage, a large representative sub-region
was extracted from the image, centred specifically on a cluster of
dislocations identified during the coarse scan. This localised
calibration region was then repeatedly processed at candidate tile
sizes spanning $0.8\,t_{\mathrm{coarse}}$ to
$3.0\,t_{\mathrm{coarse}}$ in increments of
$0.2\,t_{\mathrm{coarse}}$. For each candidate $t_i$, the detection
count $c(t_i)$ was recorded. The resulting count--size curve was
fitted to a gaussian:

\begin{equation}
c(t) = A\,\exp\!\left(-\frac{(t-\mu)^{2}}{2\sigma^{2}}\right) + C,
\end{equation}
and the optimal tile size taken as $\mu$ (rounded to the nearest
integer). The peak $\mu$ corresponds to the apparent scale at which
the trained model is most prolific, balancing the recall of faint or
small features against the dilution of spatial context in dense
regions. Calibration required 5--15\,s per image.

\textit{Inference and post-processing.} Full images were then tiled at
the calibrated size with overlapping windows. Per-tile predictions
were transformed to global image coordinates, and duplicate detections
at tile boundaries were merged by non-maximum suppression
(NMS)~\cite{neubeck2006nms} with an IoU threshold of 0.30. A final
confidence threshold of 0.025 was applied, calibrated on a subsection
of the main benchmark image and used unchanged for all images. The
calibration of this threshold, and its substantial deviation from the
validation-optimal value, is examined in
Table~\ref{tab:threshold_calibration}. Total processing time was
6--18\,s per image for the majority of the test set (calibration
5--15\,s, inference 1--3\,s), rising to $\approx 28$\,s (18\,s
calibration, 10\,s inference) for the largest benchmark image,
consistent with runtime scaling linearly with image area; processing
time was independent of dislocation density. No per-image parameter
tuning was required.

%%%%%%%%%%%%%%%%%%%%%%%%%%%%%%%%%%%%%%%%%%%%%%%%%%%%%%%%%%%%%%%%%
\section{Results and Discussion}
\label{sec:results}
%%%%%%%%%%%%%%%%%%%%%%%%%%%%%%%%%%%%%%%%%%%%%%%%%%%%%%%%%%%%%%%%%

\subsection{Method 1: Rule-Based Detection}
\label{sec:results_m1}

\begin{figure}[htb]
\centering
\begin{tikzpicture}
    \node[anchor=south west, inner sep=0] (image) at (0,0) {
        \includegraphics[width=0.48\textwidth]{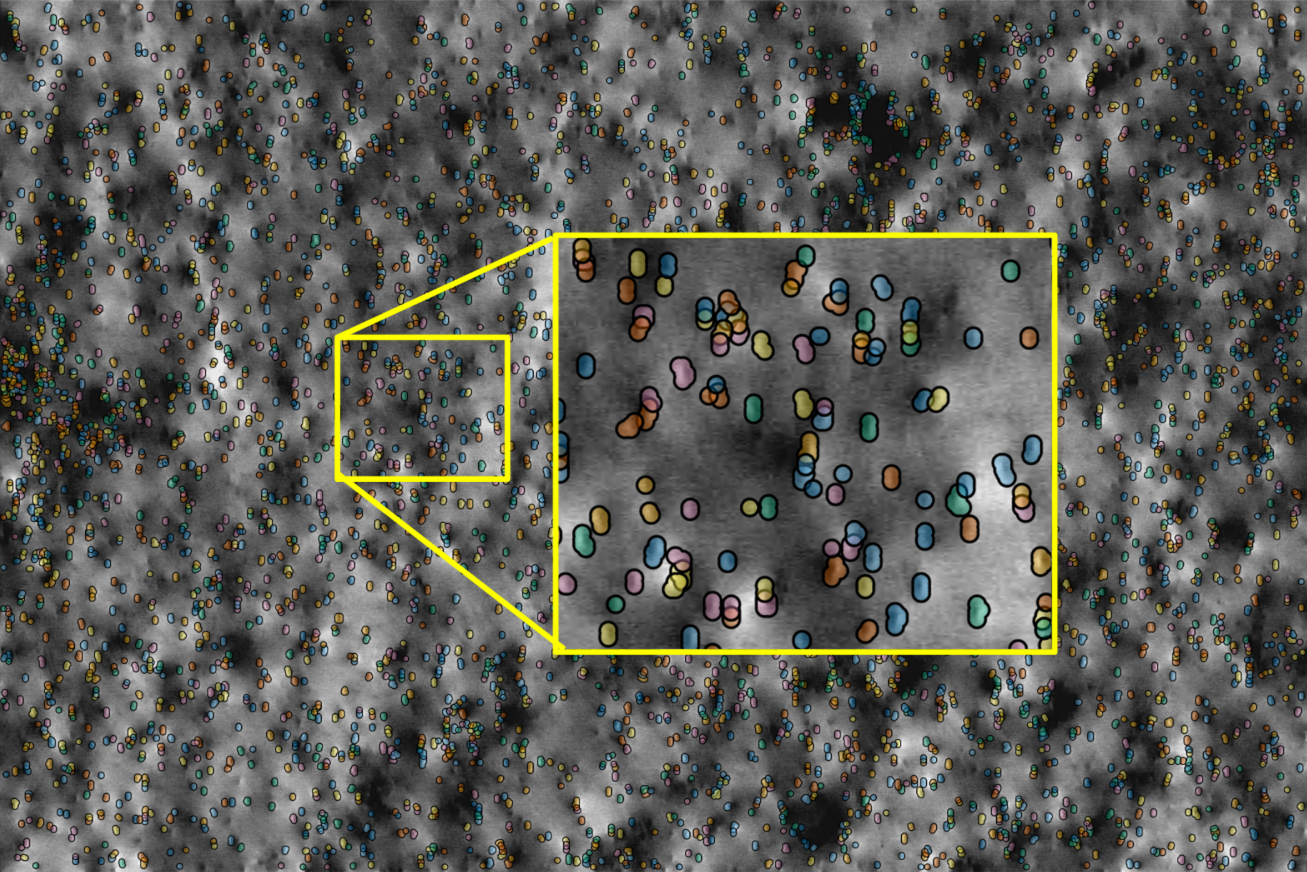}
    };
    \begin{scope}[x={(image.south east)},y={(image.north west)}]
        \draw[white, line width=1.5pt, |-|] (0.005, 0.05) -- (0.25, 0.05) 
            node[midway, above, text=white, font=\sffamily\small] {$10\,\mu\mathrm{m}$};
    \end{scope}
\end{tikzpicture}
\caption{\label{fig:m1_full} Final output of the rule-based pipeline
  on the benchmark image: 4682 clusters outlined in black and filled
  with random colours for visual separation, following lower-biased
  multi-level Otsu segmentation, per-region contrast detection, and
  mean-shift clustering. In isolated regions the clustering performs
  well, one compact cluster per dislocation, but in dense regions
  multiple neighbouring dislocations collapse into a single merged
  cluster, preventing reliable quantitative use.}
\end{figure}

\begin{figure}[h!]
\centering
\begin{subfigure}[t]{0.4\linewidth}
    \begin{tikzpicture}
        \node[anchor=south west, inner sep=0] (image) at (0,0) {
            \includegraphics[width=\linewidth, height=0.13\textheight, keepaspectratio]{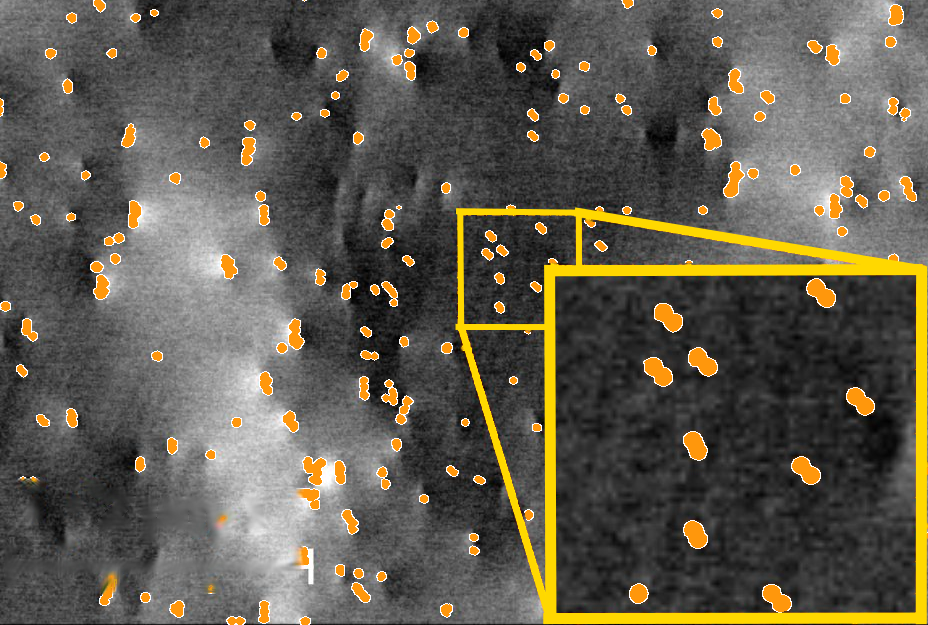}
        };
        \begin{scope}[x={(image.south east)},y={(image.north west)}]
            \draw[white, line width=1.5pt, |-|] (0.01, 0.08) -- (0.4, 0.08) 
                node[midway, above, text=white, font=\sffamily\scriptsize] {$5\,\mu\mathrm{m}$};
        \end{scope}
    \end{tikzpicture}
    \caption{}
    \label{fig:m1_pipeline_a}
\end{subfigure}
\hfill
\begin{subfigure}[t]{0.4\linewidth}
    \begin{tikzpicture}
        \node[anchor=south west, inner sep=0] (image) at (0,0) {
            \includegraphics[width=\linewidth, height=0.13\textheight, keepaspectratio]{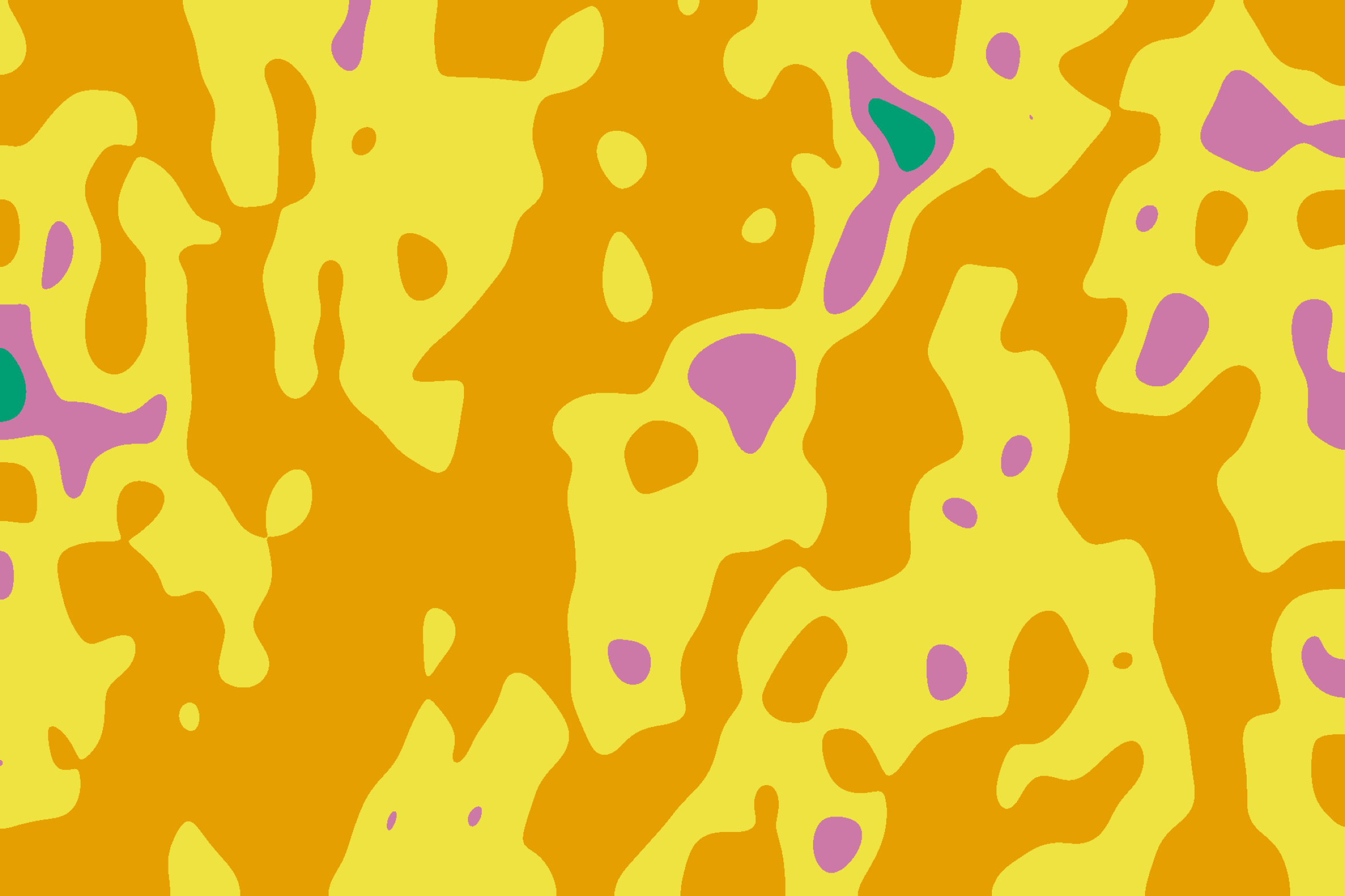}
        };
        \begin{scope}[x={(image.south east)},y={(image.north west)}]
            \draw[white, line width=1.5pt, |-|] (0.005, 0.05) -- (0.25, 0.05) 
                node[midway, above, text=white, font=\sffamily\scriptsize] {$10\,\mu\mathrm{m}$};
        \end{scope}
    \end{tikzpicture}
    \caption{}
    \label{fig:m1_pipeline_b}
\end{subfigure}
\begin{subfigure}[t]{0.75\linewidth}
    \begin{tikzpicture}
        \node[anchor=south west, inner sep=0] (image) at (0,0) {
            \includegraphics[width=\linewidth, height=0.27\textheight, keepaspectratio]{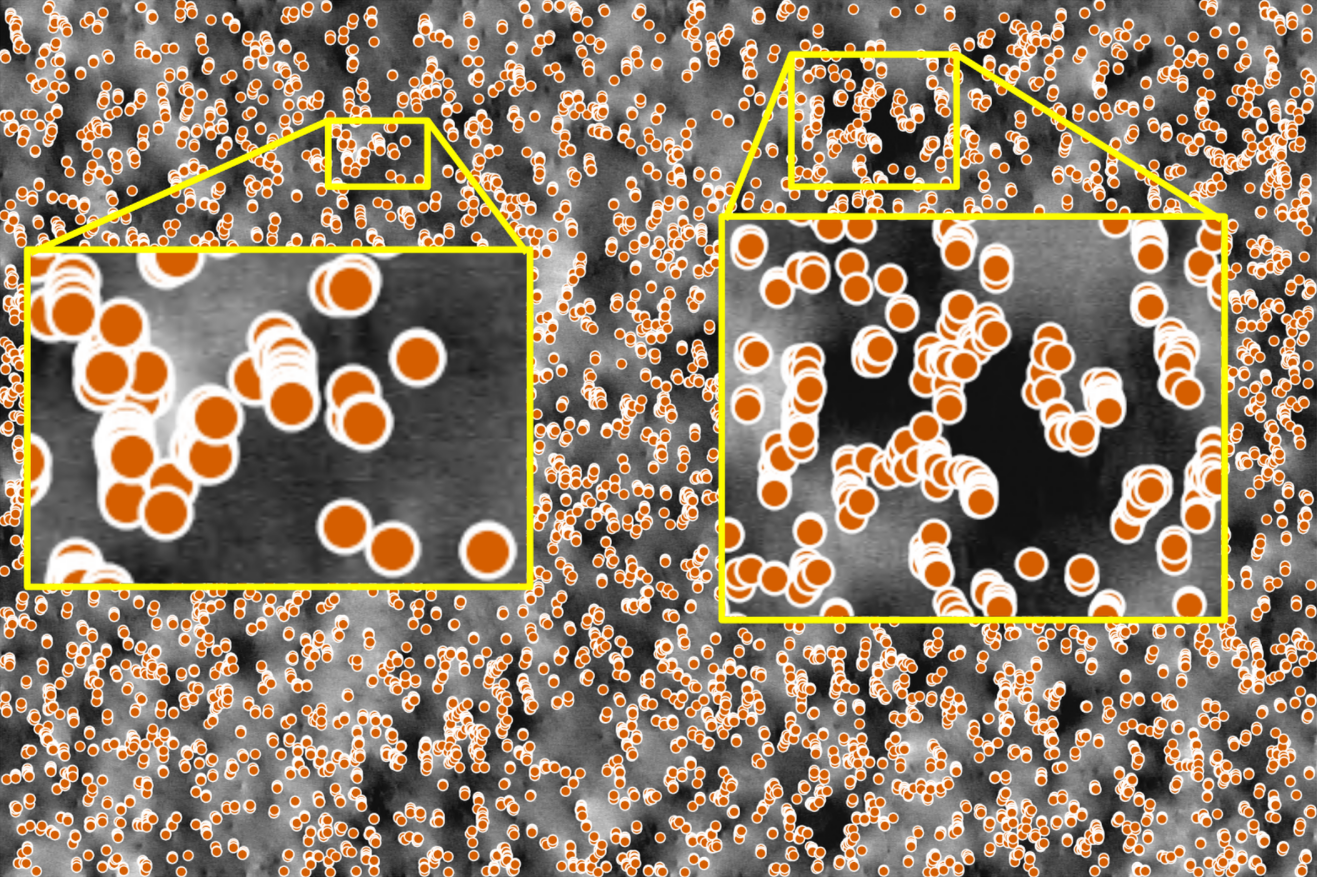}
        };
        \begin{scope}[x={(image.south east)},y={(image.north west)}]
            \draw[white, line width=1.5pt, |-|] (0.005, 0.05) -- (0.25, 0.05) 
                node[midway, above, text=white, font=\sffamily\scriptsize] {$10\,\mu\mathrm{m}$};
        \end{scope}
    \end{tikzpicture}
    \caption{}
    \label{fig:m1_pipeline_c}
\end{subfigure}
\begin{subfigure}[t]{0.4\linewidth}
    \begin{tikzpicture}
        \node[anchor=south west, inner sep=0] (image) at (0,0) {
            \includegraphics[width=\linewidth, height=0.13\textheight, keepaspectratio]{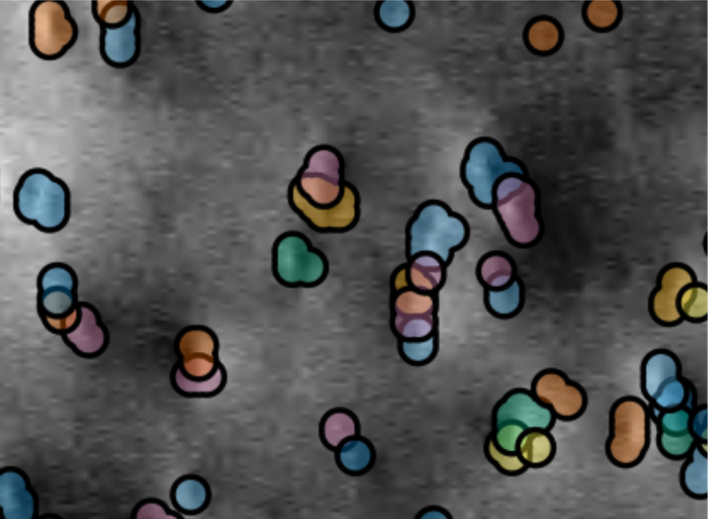}
        };
        \begin{scope}[x={(image.south east)},y={(image.north west)}]
            \draw[white, line width=1.5pt, |-|] (0.1, 0.08) -- (0.3, 0.08) 
                node[midway, above, text=white, font=\sffamily\scriptsize] {$1\,\mu\mathrm{m}$};
        \end{scope}
    \end{tikzpicture}
    \caption{}
    \label{fig:m1_pipeline_d}
\end{subfigure}
\hfill
\begin{subfigure}[t]{0.4\linewidth}
    \begin{tikzpicture}
        \node[anchor=south west, inner sep=0] (image) at (0,0) {
            \includegraphics[width=\linewidth, height=0.13\textheight, keepaspectratio]{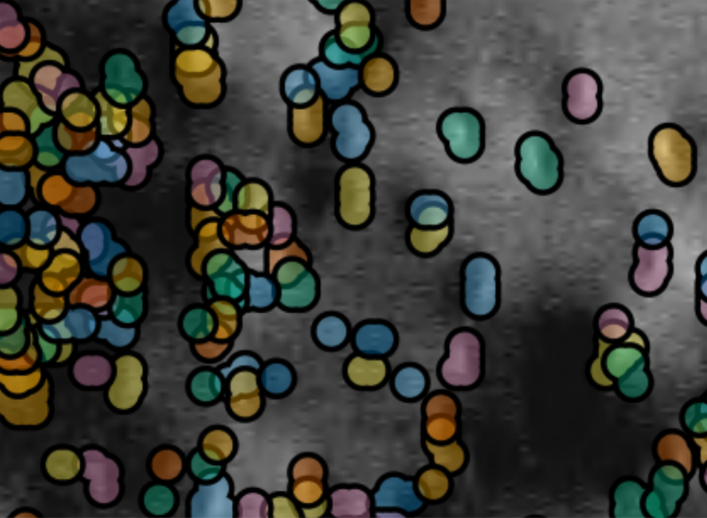}
        };
        \begin{scope}[x={(image.south east)},y={(image.north west)}]
            \draw[white, line width=1.5pt, |-|] (0.1, 0.08) -- (0.3, 0.08) 
                node[midway, above, text=white, font=\sffamily\scriptsize] {$1\,\mu\mathrm{m}$};
        \end{scope}
    \end{tikzpicture}
    \caption{}
    \label{fig:m1_pipeline_e}
\end{subfigure}
\caption{\label{fig:m1_pipeline} Stages of the rule-based
  pipeline. (a) A single contrast threshold produces dots throughout
  the image, with darker regions filled with false positives from
  background variations rather than from real dislocations, and missed
  dislocations in the brighter region. (b) Lower-biased multi-level
  Otsu segmentation splits the image into four classes, with
  thresholds concentrated in the darker areas where contrast
  sensitivity is highest. (c) With per-region contrast thresholds in
  place, dots track mostly real dislocations across the full intensity
  range. (d) Close-up of a medium-background intensity region:
  mean-shift clustering on the dots in (c) mostly works, with one
  compact cluster per dislocation in most cases, but still
  over-clusters where adjacent dislocations sit close together. (e)
  Close-up of a dark region (left hand side): even after segmentation,
  the initial dot detection picks up too many false positives from
  background noise, which carry through to clustering as extra
  clusters.}
\end{figure}
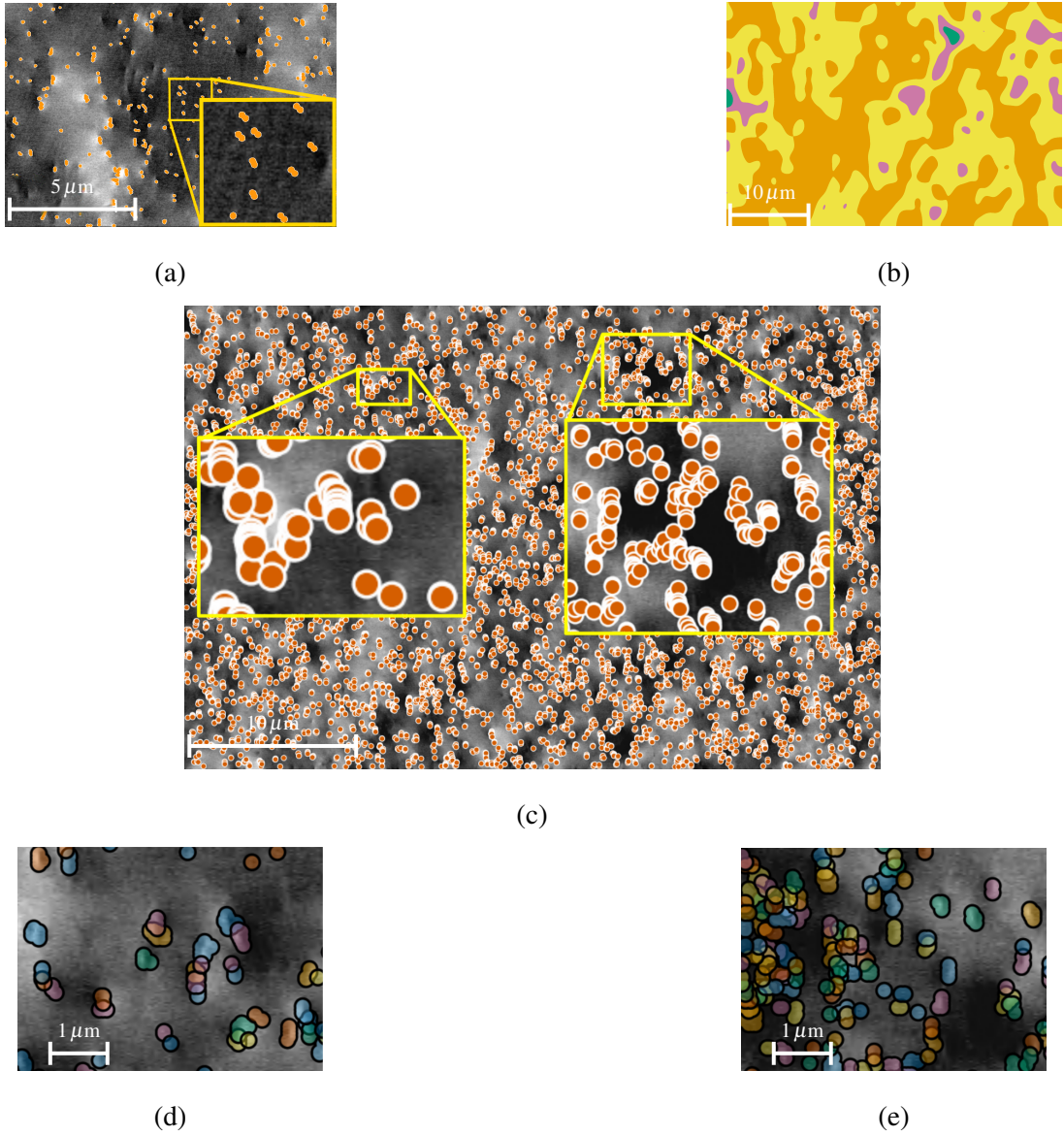

The rule-based pipeline produced 4682 candidate clusters on the
benchmark image (Fig.~\ref{fig:m1_full}), but the detections were
unreliable. This is $\approx 23\%$ higher than the manual ground
truth, but the two are only this close by chance: background noise in
dark regions adds extra clusters that push the count up, while merging
in dense regions pulls it back down. Neither the total nor the counts
behind it can be used quantitatively. Because the method required
extensive per-image manual tuning and still failed consistently in
dense or dark regions, a meaningful evaluation of precision and recall
across multiple images was infeasible. Fig.~\ref{fig:m1_pipeline}
walks through each stage of the pipeline and the fundamental
limitations which argued against further development of this approach.

A single contrast threshold across the whole image did not work
because the ECCI background intensity varies too much: raw candidate
detections (dots) fill the image, with dark regions producing false
positives from background variations rather than from real
dislocations (Fig.~\ref{fig:m1_pipeline}a). Lower-biased multi-level
Otsu segmentation~\cite{otsu1979threshold} solves this by partitioning
the image into intensity classes with thresholds concentrated in the
darker parts of the image, where contrast sensitivity is highest
(Fig.~\ref{fig:m1_pipeline}b). With per-region contrast thresholds in
place (typically from 10 out of 255 in the darkest regions up to 100
in moderately bright grains), the dots track mostly real dislocations
across the full intensity range, with several dots around each dipole
rather than covering the darker regions
(Fig.~\ref{fig:m1_pipeline}c). However, this stage alone required 5--8
manually-tuned parameters per image (contrast thresholds for each
segmented region, clustering bandwidth, window size), with 15--30
minutes of trial-and-error tuning per image. The mean-shift clustering
step that follows works well in isolated regions, producing one
compact cluster per dislocation, but fails in dense regions: any
bandwidth that merges multiple dots within a single dislocation also
merges adjacent dislocations into one cluster
(Fig.~\ref{fig:m1_pipeline}d). No single parameter set could both
merge dots within a dislocation and separate adjacent
dislocations. Very dark regions remain difficult even after
segmentation, contrast there sits close to the noise floor, making
faint dislocations hard to separate from background noise
(Fig.~\ref{fig:m1_pipeline}e).

Several of these issues could likely be reduced with further work,
better clustering, dedicated noise rejection in dark regions, or
adaptive parameter selection, but the core problem is that the
clustering step has to use one bandwidth to both merge dots within a
dislocation and separate adjacent dislocations, and refining the other
steps does not solve this issue. Combined with the per-image tuning
burden and overall slow dot rasterising and clustering times, further
development of the approach was not judged to be worthwhile. It is
excluded from quantitative comparison, but its limitations directly
motivated the learning-based approaches that follow.

\subsection{Method 2: Multi-Stage CNN}
\label{sec:results_m2}

\begin{figure}[htb]
\centering
\begin{tikzpicture}
    \node[anchor=south west, inner sep=0] (image) at (0,0) {
        \includegraphics[width=0.48\textwidth]{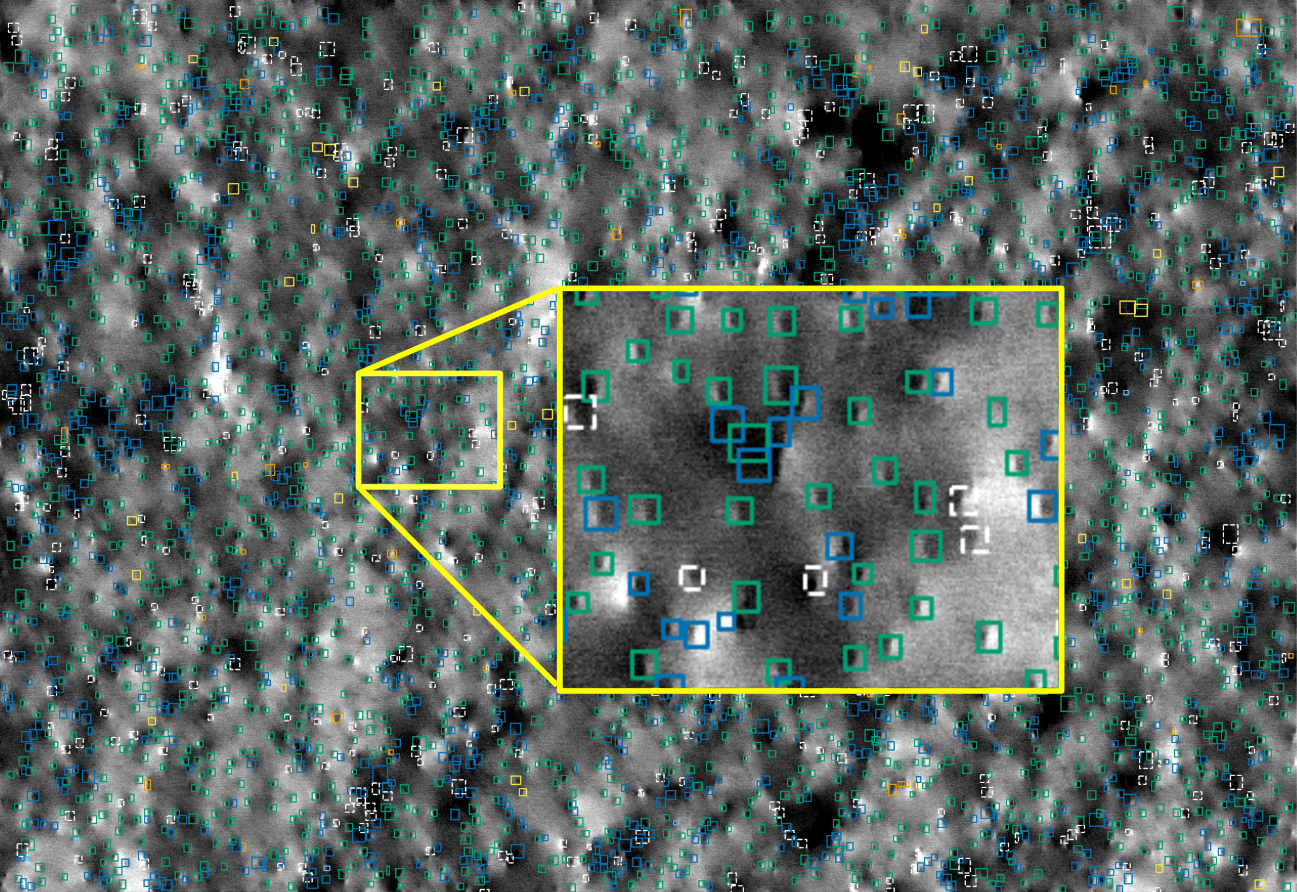}
    };
    \begin{scope}[x={(image.south east)},y={(image.north west)}]
        \draw[white, line width=1.5pt, |-|] (0.005, 0.05) -- (0.25, 0.05)
            node[midway, above, text=white, font=\sffamily\small] {$10\,\mu\mathrm{m}$};
    \end{scope}
\end{tikzpicture}
\caption{\label{fig:m2_full} Final output of the multi-stage CNN
  pipeline on the benchmark image: 3312 detections shown as
  colour-coded bounding boxes by pipeline branch (green = single,
  yellow = faint/ultra-faint, blue = multiple close by, orange =
  tri-pole, cyan = quad-pole), with white dotted boxes marking
  \textit{unknown} regions that are flagged for manual review rather
  than committed to as detections.}
\end{figure}

The multi-stage CNN detected 3312 dislocations on the benchmark image
(Fig.~\ref{fig:m2_full}), giving a counting accuracy of $87.0 \pm
1.1\%$ relative to the manual ground truth of $3805 \pm 46$. Manual
verification of these detections identified 3195 true positives and
117 false positives ($3.5\%$ of detections). Because the number of
false negatives is calculated relative to the statistical ground
truth, it inherits the same variance, yielding $610 \pm 46$ missed
dislocations ($16.0\%$ of the ground truth), of which 605 were
individually located in the verification pass
(Table~\ref{tab:method2_errors}). This corresponds to a precision of
$96.5\%$, a recall of $84.1\%$, and an F1 score of
$89.8\%$. Processing time was $\approx 10$~hours 20~minutes, with cost
growing super-linearly in dense regions due to iterative re-centring
and neighbour searches.

Table~\ref{tab:method2_errors} breaks down the errors. False negatives
are split between those the pipeline flagged for review
(\emph{Unknown} regions) and those it missed silently. This
distinction matters because the two have very different operational
consequences, a flagged miss costs minimal review time, whilst a
silent miss escapes detection entirely or requires a full image review
(effectively doing the whole process manually again).

\begin{table}[htb]
\centering
\caption{\label{tab:method2_errors} Error breakdown for the
  multi-stage CNN on the benchmark image
  (Fig.~\ref{fig:ecci_example_with_zoom_int_single_dislocation}). Ground
  truth: $3805 \pm 46$ dislocations (95\% CI, $n=3$). The breakdown
  lists the 605 misses individually located in the verification pass;
  the statistical false-negative total is $610 \pm 46$. False
  negatives are divided into two distinct categories:
  \emph{system-flagged} misses, which fall inside \emph{Unknown}
  regions that the pipeline explicitly raised for manual review, and
  \emph{silent} misses, which escaped detection entirely.}
\begin{ruledtabular}
\begin{tabular}{lcc}
\textbf{Error type} & \textbf{Count} & \textbf{\% of category} \\
\hline
\multicolumn{3}{l}{\emph{False positives (Total 117)}} \\
\quad Double-counts                            & 85  & 72.6 \\
\quad Solitary false positives                 & 22  & 18.8 \\
\quad Inside \emph{Unknown}                    & 10  & 8.5 \\
\hline
\multicolumn{3}{l}{\emph{False negatives (Total = 605)}} \\
\quad System-flagged (inside \emph{Unknown})   & 397 & 65.6 \\
\quad Silent --- perimeter blind-zone          & 60  & 9.9 \\
\quad Silent --- dense / ambiguous             & 115 & 19.0 \\
\quad Silent --- very faint                    & 33  & 5.5  \\
\hline
\multicolumn{3}{l}{\emph{Summary statistics}} \\
\quad True positives (TP)                      & 3195 & --- \\
\quad Counting accuracy                        & $87.0 \pm 1.1\%$ & --- \\
\quad Precision                                & $96.5\%$ & --- \\
\quad Recall                                   & $84.1\%$ & --- \\
\quad F1 score                                 & $89.8\%$ & --- \\
\end{tabular}
\end{ruledtabular}
\end{table}

\begin{figure}[h!]
\centering

\begin{subfigure}[t]{0.432\linewidth}
    \includegraphics[width=\linewidth, clip]{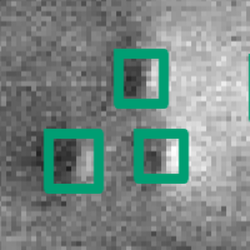}
    \caption{}
    \label{fig:m2_failures_a}
\end{subfigure}
\hfill
\begin{subfigure}[t]{0.432\linewidth}
    \includegraphics[width=\linewidth, clip]{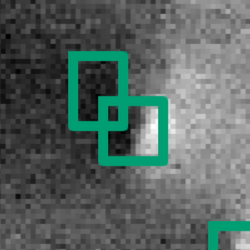}
    \caption{}
    \label{fig:m2_failures_b}
\end{subfigure}

\vspace{0.5em}

\begin{subfigure}[t]{0.432\linewidth}
    \includegraphics[width=\linewidth, clip]{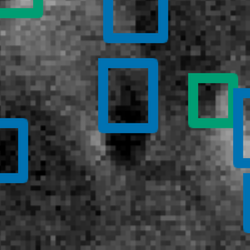}
    \caption{}
    \label{fig:m2_failures_c}
\end{subfigure}
\hfill
\begin{subfigure}[t]{0.432\linewidth}
    \begin{tikzpicture}
        \node[anchor=south west, inner sep=0] (image) at (0,0) {
            \includegraphics[width=\linewidth, clip]{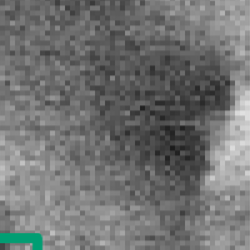}
        };
        \begin{scope}[x={(image.south east)},y={(image.north west)}]
            \draw[white, line width=1.5pt, |-|] (0.1, 0.08) -- (0.9, 0.08) 
                node[midway, above, text=white, font=\sffamily\small] {$1\,\mu\mathrm{m}$};
        \end{scope}
    \end{tikzpicture}
    \caption{}
    \label{fig:m2_failures_d}
\end{subfigure}

\caption{\label{fig:m2_failures} Representative failure modes of the
  multi-stage CNN. (a) Clean detections in a moderate-density region:
  one box per dislocation, well-centred. (b) A double-count: a single
  dislocation receives two bounding boxes from neighbouring scan
  positions because the first box was off-centre and left enough
  residual contrast for a second detection. (c) An off-target box: the
  locator places a box that overlaps the dislocation enough to count
  as a true positive but is not centred on it. (d) Perimeter
  blind-zone: dislocations near the image edge (left side) are missed
  because the training set patches only labelled dislocations within
  the central quarter as positives, so the classifier never learned to
  flag them.}
\end{figure}

Fig.~\ref{fig:m2_failures} outlines the main failure modes with
Fig.~\ref{fig:m2_failures_a} showing correctly identified
dislocations. Two failure modes dominate. First, 85 of the 117 false
positives ($73\%$) were double-counts of dislocations
(Fig.~\ref{fig:m2_failures}b). The locator placed boxes that
overlapped the dislocation enough to register as true positives but
were off-centre on harder patches
(Fig.~\ref{fig:m2_failures}c). Off-centre boxes left enough residual
contrast for the scanner to fire again on the neighbouring patch, and
the deduplication step did not recognise these slight offsets as the
same feature. The same off-centre placement also weakens the
precision: many true positives are boxes whose overlap is enough to
pass a lenient IoU threshold but not where a human annotator would
draw the box, overstating localisation quality. Several design choices
plausibly contribute to the locator's off-centre behaviour, MSE on box
coordinates rather than overlap, no in-model validity constraints on
the four predicted corners, and a limited ($\approx 1000$) labelled
set biased toward already-centred dislocations, but we did not isolate
which of these dominated.

Second, the \emph{Unknown}-flagging mechanism caught 397 of 605 false
negatives ($66\%$), mostly dense clusters and locator convergence
failures, flagging them for manual review rather than committing to a
low-confidence detection. The remaining 208 misses were silent: 60
near image perimeters (Fig.~\ref{fig:m2_failures}d), missed because
the training set patches only labelled dislocations within the central
quarter as positives; 115 from dense or ambiguous features the
pipeline neither detected nor flagged; and 33 from very faint
dislocations whose contrast fell below what the classifier learned to
respond to. \emph{Unknown} therefore caught just over two in three
misses, however very few of the double-counts, instead it highlights
tricky areas which it was unable to classify with high confidence, not
the localisation errors that produced the dominant false-positive
mode.

Each failure has a targeted fix: overlap-based deduplication,
retraining the locator on a larger and more varied label set,
relabelling perimeter patches as positives, an improved dense-region
handler; but the underlying problem is structural. The pipeline uses
three independently-trained networks coordinated by hand-crafted
logic, with each stage solving a task for the downstream
one. Stage-level fixes tend to expose new failure modes at the next
interface. Furthermore, the sequential sliding-window, centring and
local search approach resulted in massive runtimes (approximately 10
hours 20 minutes on the benchmark image). While this could be improved
through parallelisation---either by batch-processing multiple patches
simultaneously to fully saturate a single GPU's compute capacity, or
by distributing independent image tiles across multiple discrete
GPUs---the runtimes would likely remain on the order of tens of
minutes per image. Combined with the pipeline's super-linear
computational scaling in dense regions, these structural limitations
directly motivated the move to a unified single-stage detector trained
end-to-end for one highly efficient pass over the image.

\subsection{Method 3: YOLOv8 with Adaptive Tile Sizing}
\label{sec:results_m3}

\subsubsection{Adaptive tile-size calibration}
\label{sec:results_tile}

\begin{figure}[htb]
\centering
\begin{subfigure}[t]{0.89\linewidth}
    \includegraphics[width=\linewidth]{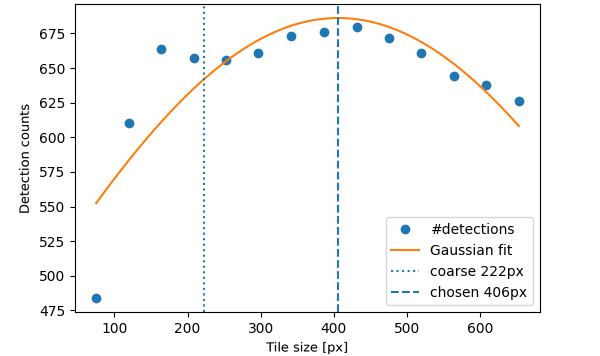}
    \caption{}
    \label{fig:tile_calibration_a}
\end{subfigure}
\hfill
\begin{subfigure}[t]{0.89\linewidth}
    \includegraphics[width=\linewidth]{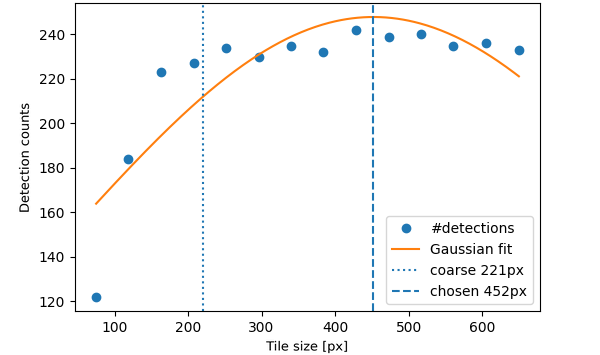}
    \caption{}
    \label{fig:tile_calibration_b}
\end{subfigure}
\caption{\label{fig:tile_calibration} Adaptive tile-size
  calibration. Detection count (blue points) versus candidate tile
  size for two separate smaller ECCI images, with the initial coarse
  estimate (dotted line), fitted gaussian (orange), and selected tile
  size $\mu$ (dashed line). (a) exhibits a primary peak alongside a
  secondary maximum, while (b) rises to a flat-topped plateau. The
  curves are noisy and not strictly gaussian, but each contains a
  clear high-yield region whose centre is well identified by $\mu$.}
\end{figure}

Apparent dislocation size in ECCI varies with magnification, and
YOLOv8 works best for features near a model-preferred apparent
size.~\cite{Lin_2017_CVPR} The adaptive tile-sizing procedure
(Sec.~\ref{sec:methods_yolo}) crops each image into separate sections
so dislocations fall within this preferred
range. Fig.~\ref{fig:tile_calibration} shows representative
count-versus-tile-size curves for two separate ECCI micrographs. The
curves are noisy, and Fig.~\ref{fig:tile_calibration_a} has a primary
peak alongside a secondary maximum, while
Fig.~\ref{fig:tile_calibration_b} rises to a flat-topped plateau, but
each contains a clear high-yield region. A least squares gaussian fit
is used, not because the data always strictly follows this
distribution, but because $\mu$ is more robust to asymmetric noise and
secondary peaks than the global maximum or a sharper parametric
fit. The procedure therefore identifies a workable rather than ideal
tile size; the flat top in Fig.~\ref{fig:tile_calibration_b} (count
varying by $\approx 10$ around a peak of $\approx 240$) shows
performance can be insensitive to moderate deviations in scale. Across
the test set, the procedure converged without manual intervention,
adding 5--15~s per image (18~s for the largest benchmark image).

\subsubsection{Confidence-threshold calibration}
\label{sec:results_conf}

\begin{figure}[htb]
\centering
\includegraphics[width=0.9\linewidth]{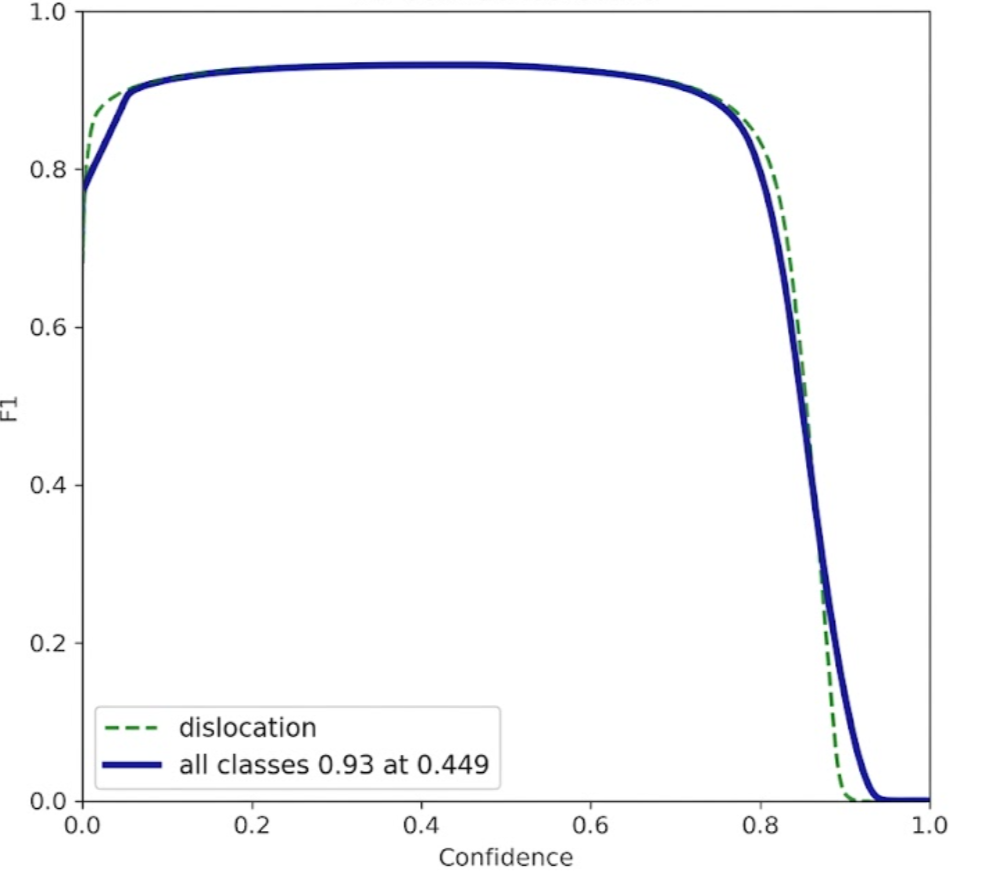}
\caption{\label{fig:f1_conf} Validation F1 versus confidence threshold
  for the trained YOLOv8 model (dislocations class). The curve
  plateaus between $\mathrm{confidence} \approx 0.05$ and $0.7$ and
  offers no sharp preferred threshold within this range; the
  conventional default of $0.25$ sits inside the plateau. F1 declines
  gradually below $\mathrm{confidence} \approx 0.05$ and steeply below
  $0.025$, the region in which deployment counting accuracy is in fact
  maximised (Table~\ref{tab:threshold_calibration}). The bold summary
  line is the Ultralytics-default smoothed average across
  classes;~\cite{yolov8_ultralytics} for this single-class model it is
  the same data convolved with a smoothing filter.}
\end{figure}

The conventional procedure for setting YOLO's deployment confidence
threshold relies on the validation F1--confidence curve
(Fig.~\ref{fig:f1_conf}, solid blue line).~\cite{yolov8_ultralytics}
The F1 score is the harmonic mean of precision and recall, and the
curve traces this score over the validation set as a function of the
minimum confidence required for a detection to be retained. The
deployment threshold is conventionally chosen where F1 is
maximised. For the trained model, F1 plateaus between a confidence of
$\approx 0.05$ and $0.7$ at F1 $\approx 0.93$, with the framework
default of $0.25$ sitting within this plateau. Above a confidence of
$0.7$ the score collapses, standard high-threshold behaviour. Below a
confidence of $\approx 0.05$ the curve declines gradually before
dropping sharply below $0.025$ to F1 $\approx 0.7$ at the origin. The
validation curve therefore offers no sharp preferred operating point
within the plateau and a strong reason to avoid the steep region below
it.

Threshold calibration was performed on a representative subsection of
the main benchmark image. The ground truth for this subsection was
established via careful manual counting, yielding 599 dislocations.

Deployment behaviour on the calibration subsection disagrees sharply
with this picture (Table~\ref{tab:threshold_calibration}). Counting
accuracy increases as the threshold is lowered from a confidence of
$0.25$ to $0.025$, a gain of $8.5\%$ ($90.8\%$ to $99.3\%$), whilst
precision remains robust at or above $96.8\%$. Although the absolute
F1 score peaks fractionally higher at a confidence of $0.05$
($96.6\%$), the score at $0.025$ is almost identical ($96.5\%$). The
counting accuracy optimum therefore sits around a confidence of
$0.025$, not in the validation plateau, but at the upper edge of the
steep decline that conventional validation would treat as
failure. While reducing the threshold further to $0.015$ yields a raw
count marginally closer to the ground truth ($99.5\%$ accuracy), it
does so by accumulating false positives rather than true detections
(only finding 1 more TP), which drives the F1 score down to
$96.1\%$. The $0.025$ threshold avoids too much false-positive
inflation, and it was found to generalise well across the varying
dislocation densities and contrast conditions of the wider multi-image
test set.

\begin{table}[htb]
\centering
\caption{\label{tab:threshold_calibration} YOLO performance on the
  calibration subsection across confidence thresholds ($\mathrm{IoU} =
  0.30$). Manual ground truth: 599 dislocations.}
\begin{ruledtabular}
\begin{tabular}{lccccccc}
\textbf{Conf.} & \textbf{Detections} & \textbf{TP} & \textbf{FP} & \textbf{FN} & \textbf{Prec. (\%)} & \textbf{Recall (\%)} & \textbf{F1 (\%)} \\
\hline
0.015 & 602 & 577 & 25 & 22 & 95.8 & 96.3 & 96.1 \\
0.025 & 595 & 576 & 19 & 23 & 96.8 & 96.2 & 96.5 \\
0.05  & 589 & 574 & 15 & 25 & 97.5 & 95.8 & 96.6 \\
0.10  & 566 & 560 & 6  & 39 & 98.9 & 93.5 & 96.1 \\
0.25  & 544 & 544 & 0  & 55 & 100.0 & 90.8 & 95.2 \\
\end{tabular}
\end{ruledtabular}
\end{table}

This breaks the typical pattern for image object detection, where low
confidence reflects classification ambiguity and more aggressive
filtering improves precision. Two domain-specific factors explain why
ECCI behaves differently. First, classification is rarely ambiguous:
the dipole contrast is morphologically distinct from grain boundaries,
surface steps, and noise, and the training data contain few
dislocation-mimicking features. Second, signal strength varies
enormously with channelling conditions, grain orientation, dislocation
character, and local background brightness. The confidence score
therefore appears to primarily encode \emph{signal strength} (``I see
a faint dislocation'') rather than \emph{classification uncertainty}
(``I am unsure whether this is a dislocation''). The validation set
inherits the same class separation as the training data, so the F1
curve correctly identifies these low-confidence detections as
uncertain. It also happens that in this setting low confidence is
informative about visibility rather than category, and most of those
uncertain detections are real. We frame this as an application domain
observation, rather than a general critique of validation-based
threshold selection: datasets with more confusable false positives
would behave differently. The threshold selected on the calibration
subsection ($0.025$) was deployed without adjustment on all test
images, achieving a counting accuracy of $98.6 \pm 1.2\%$ on the
benchmark image (Sec.~\ref{sec:results_m3_errors}) and a precision of
$98.7\%$ with a recall of $98.7\%$ across the aggregate test set
(Sec.~\ref{sec:results_multi}), indicating that the calibration
transfers across images and dislocation densities.

\subsubsection{Error breakdown}
\label{sec:results_m3_errors}

\begin{figure}[htb]
\centering
\begin{subfigure}[t]{0.99\linewidth}
    \begin{tikzpicture}
        \node[anchor=south west, inner sep=0] (image) at (0,0) {
            \includegraphics[width=\linewidth, height=0.35\textheight, keepaspectratio]{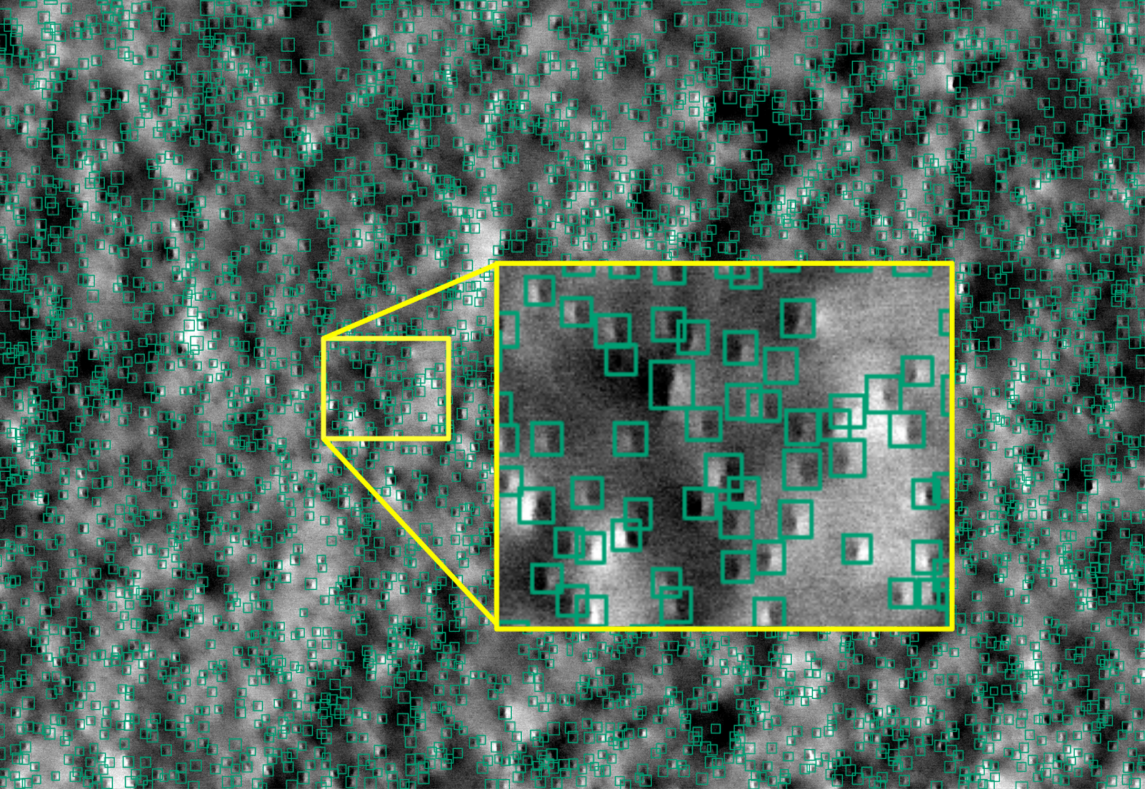}
        };
        \begin{scope}[x={(image.south east)},y={(image.north west)}]
            \draw[white, line width=1.5pt, |-|] (0.005, 0.05) -- (0.25, 0.05) 
                node[midway, above, text=white, font=\sffamily\small] {$10\,\mu\mathrm{m}$};
        \end{scope}
    \end{tikzpicture}
    \caption{}
    \label{fig:yolo_raw}
\end{subfigure}

\vspace{0.5em}

\begin{subfigure}[t]{0.99\linewidth}
    \begin{tikzpicture}
        \node[anchor=south west, inner sep=0] (image) at (0,0) {
            \includegraphics[width=\linewidth, height=0.35\textheight, keepaspectratio]{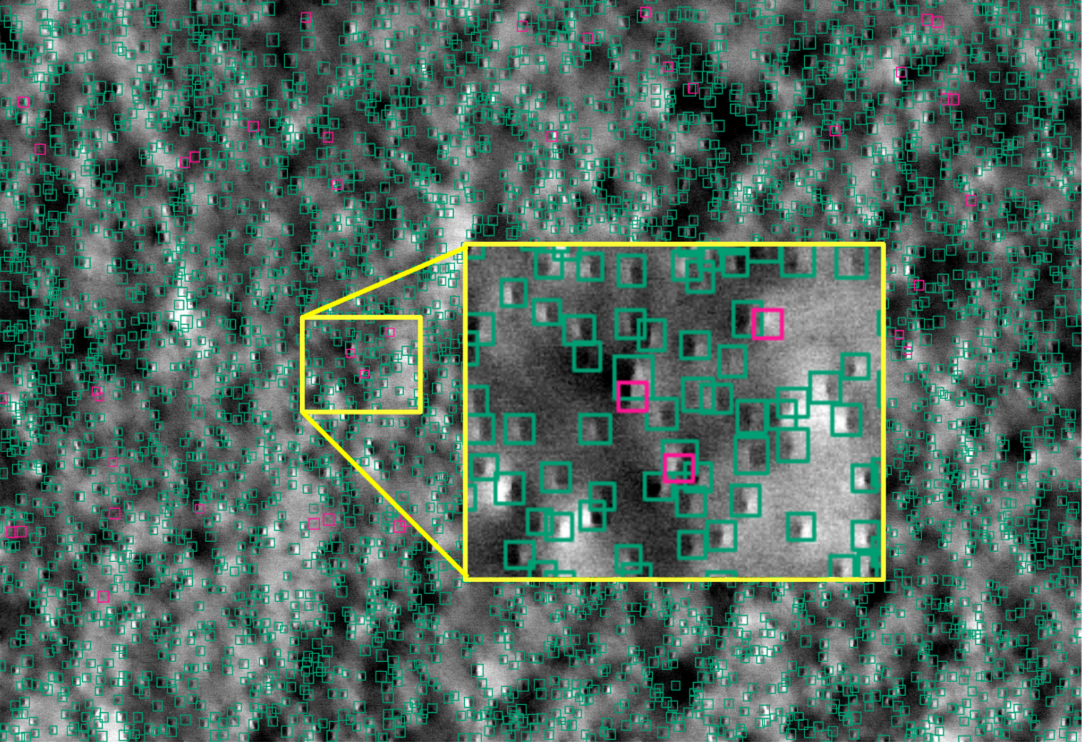}
        };
        \begin{scope}[x={(image.south east)},y={(image.north west)}]
            \draw[white, line width=1.5pt, |-|] (0.005, 0.05) -- (0.25, 0.05) 
                node[midway, above, text=white, font=\sffamily\small] {$10\,\mu\mathrm{m}$};
        \end{scope}
    \end{tikzpicture}
    \caption{}
    \label{fig:yolo_corrected}
\end{subfigure}
\caption{\label{fig:yolo_detection} YOLO detection on the benchmark
  image at $\mathrm{confidence} = 0.025$, $\mathrm{IoU} = 0.30$. (a)
  Raw output: 3858 detections (green). (b) Manual correction: false
  negatives added (magenta). The 116 total corrections (74 false
  positives plus 42 false negatives) represent $3.0\%$ of the ground
  truth.}
\end{figure}

Table~\ref{tab:yolo_errors} reports the error count. A pass at a
confidence of $0.025$ was run and each dislocation was verified
manually. This resulted in 74 false positives and 42 false negatives,
giving a corrected total of 3826, which lies within the 95\%
confidence interval of the ground truth ($3805 \pm 46$); the raw count
of 3858 overshoots the mean by $1.4\%$, reflecting the same
intra-operator variation that produces the ground-truth
uncertainty. Fig.~\ref{fig:yolo_detection} shows the corresponding
detections with manual corrections overlaid;
Table~\ref{tab:yolo_errors} breaks the errors down by type.

\begin{table}[htb]
\centering
\caption{\label{tab:yolo_errors} Error breakdown for YOLO on the
  benchmark image at a confidence of $0.025$, $\mathrm{IoU} = 0.30$,
  from a single detailed verification pass.}
\begin{ruledtabular}
\begin{tabular}{lcc}
\textbf{Error type} & \textbf{Count} & \textbf{\% of category} \\
\hline
\multicolumn{3}{l}{\emph{False positives (Total = 74)}} \\
\quad Ambiguous perimeter features    & 33  & 44.6 \\
\quad Ambiguous contrast features     & 31  & 41.9 \\
\quad Double-count on single feature  & 10  & 13.5 \\
\hline
\multicolumn{3}{l}{\emph{False negatives (Total = 42)}} \\
\quad Within dense clusters           & 24 & 57.1 \\
\quad Extremely faint contrast        & 18  & 42.9 \\
\end{tabular}
\end{ruledtabular}
\end{table}

The 74 false positives fall into three categories: 33 (44.6\%) are
ambiguous image perimeter features, subtle intensity rises that may or
may not be the edges of dislocations extending beyond the field of
view; a further 31 (41.9\%) are ambiguous contrast features within the
image whose interpretation remained uncertain even under careful
manual inspection but edging closer to a dislocation than not; and the
remaining 10 (13.5\%) are double-counts of single dislocations,
against 85 such errors for the multi-stage CNN. The dominant failure
mode of Method~2 is therefore strongly suppressed here.

No detections were triggered by annotations on this image, although
unfamiliar scale-bar styles or other markings absent from the training
set can occasionally be picked up, and are trivially removed by
cropping.

False negatives are split into two regimes. Just over half (24 of 42)
lie within dense clusters where dislocation spacing approaches the
feature diameter and annotation itself becomes an
interpretation. These are the same regions in which repeated manual
counts can vary the most between operators. The remaining 18 false
negatives are extremely faint dislocations near the practical limit of
the imaging conditions, only marginally distinguishable from
background noise (it is highly likely several of these are often
missed in manual counting). Neither category is the result of a model
failure: both sit at the edge of what ECCI can resolve and what a
human can annotate consistently, so the same ambiguity will be present
in the labels in the training data which has carried through to the
model. It appears that most errors in YOLOv8 come from human error in
the labelling of the dataset.

\subsection{Comparative Summary}
\label{sec:results_compare}

Table~\ref{tab:comparison} sets the two completed methods side by side
on the benchmark image. The unified single-stage detector (YOLOv8)
improves counting accuracy from $87.0\%$ to $98.6\%$, reduces the
false-positive rate from $3.5\%$ to $1.9\%$ of detections and false
negatives by a factor of $\approx 14$, and processes images
$\approx1330\times$ faster. The accuracy obtained ($98.6 \pm 1.2\%$)
corresponds to a deviation of $1.4\%$ from the mean manual count, with
the verification-corrected total falling inside the ground-truth
confidence interval, placing automated detection near the practical
limit of the measurement itself with manual counting.

\begin{table}[htb]
\centering
\caption{\label{tab:comparison} Performance comparison on the
  benchmark image. Ground truth $3805 \pm 46$ dislocations (95\% CI,
  $n = 3$). Hardware: NVIDIA RTX 3080.}
\begin{ruledtabular}
\begin{tabular}{lcc}
\textbf{Metric} & \textbf{Multi-stage CNN} & \textbf{YOLOv8} \\
\hline
Detected count           & 3312                  & 3858 \\
Counting accuracy        & $87.0 \pm 1.1\%$     & $98.6 \pm 1.2\%$ \\
Precision                & $96.5\%$             & $98.1\%$ \\
Recall                   & $84.1\%$             & $98.9\%$ \\
F1 score                 & $89.8\%$             & $98.5\%$ \\
False positives          & 117 (3.5\%)           & 74 (1.9\%) \\
False negatives (located) & 605 (15.9\%)         & 42 (1.1\%) \\
Processing time          & $\approx 620$~min       & \begin{tabular}[t]{@{}c@{}}$\approx 28$\,s \\ (18\,s calib.\ + 10\,s infer.)\end{tabular} \\
Runtime scaling          & Super-linear (density)  & Linear (area) \\
Manual parameters        & 1 per image          & 0 (automatic) \\
\end{tabular}
\end{ruledtabular}
\end{table}

Both methods were trained on similar datasets and ran on the same
hardware, so the gap is not a question of data volume or computing
power. Three architectural differences, each surfaced in the
per-method results above, are jointly sufficient to account for
it. CIoU loss optimises overlap directly where MSE on box corners
rewards coordinate proximity, which is why the 85 off-centre
double-counts that dominated Method~2's errors fall to 10 cases under
YOLO. The calibrated tiles preserve grain-boundary and neighbourhood
context that Method~2's $\approx 36$\,px patches excluded,
contributing to the improvement in both dense and faint
regions. Training the box and confidence score from a shared
representation removes the hand-crafted interfaces between Method~2's
three networks, the same interfaces that let the \emph{Unknown}
channel catch 397 of 605 missed dislocations whilst missing 85
double-counts. These three architectural choices account for the gap
on this dataset, but the comparison is not purely one of architecture:
YOLOv8 also reflects years of open-source development and refinement
that a single-group custom pipeline cannot match. Some fraction of the
measured advantage is therefore attributable to that accumulated
maturity rather than to the architectural choices in isolation.

\subsubsection{Multi-image evaluation}
\label{sec:results_multi}

While comparisons using the benchmark image give a useful baseline for
analysis and cross reference of each method's advantages and
disadvantages, and is a fast way of discarding clearly limited
techniques, testing whether YOLO generalises further requires a larger
image set. The YOLOv8 model was therefore applied to a wider test set
spanning a range of magnifications, dislocation densities, and grain
morphologies, including the benchmark image, totalling 7451
manually-counted dislocations (Table~\ref{tab:multi_image}). The model
produced 7452 raw detections, of which 7352 were true positives,
giving aggregate precision of $98.7\%$, a recall of $98.7\%$, and an
F1 score of $98.7\%$. Total required manual correction across all
images amounted to $2.7\%$ of the ground truth.

\begin{table}[htb]
\centering
\caption{\label{tab:multi_image} Aggregate YOLO performance across the
  multi-image test set, including the benchmark image (3826 of 7451
  dislocations), of which the calibration subsection is a part.}
\begin{ruledtabular}
\begin{tabular}{lc}
\textbf{Metric} & \textbf{Value} \\
\hline
Manual ground truth (total)              & 7451 \\
YOLO detections (raw)                    & 7452 \\
True positives                           & 7352 \\
False positives                          & 100 \\
False negatives                         & 99 \\
Precision                                & $98.7\%$ \\
Recall                                   & $98.7\%$ \\
F1 score                                 & $98.7\%$ \\
Required manual corrections / ground truth divergence       & $2.7\%$ \\
\end{tabular}
\end{ruledtabular}
\end{table}

The aggregate precision and recall are within $\approx 1\%$ of the
values observed for the benchmark image. Because the benchmark image
contributes roughly half the aggregate (3826 of 7451 dislocations),
generalisation is better judged from the remaining images alone, which
give a precision of $99.3\%$ and a recall of $98.4\%$, indicating that
performance generalises beyond the benchmark image. Within these
remaining images, a large fraction of the false positives came from
artefacts in the ECCI micrographs absent in the training set leading
to white and black spots misidentified as dislocations, which are a
reflection of the labelling coverage rather than the model
itself. With $2.7\%$ of ground truth requiring manual correction
across 7451 dislocations, the detector approaches the same precision
band as the manual identification itself.

\subsection{Further Improvements}

The clearest routes to reduce the error lie in improving the labelled
dataset rather than the model itself. False positives on the
multi-image set were often artefacts in the ECCI micrographs absent
from training, isolated bright or dark spots may be common in some
samples but were largely missing from the training data, and hence the
network has no class label to assign to them, so sometimes will file
them under \emph{dislocation}. Adding an explicit \emph{artefact}
class would give the model somewhere to route these features, supply
useful negative examples during training, and as a side effect give
the sample growers more information about potential foreign particles
in samples.

Another strength of YOLOv8 over more custom methods is the ease of
use: the YOLOv8 framework used here is open, well-documented, and
straightforward to train on a single consumer GPU. Other groups
working with different growth processes, magnification ranges, or
microscope conditions can therefore retrain the network on their own
images, targeting the feature densities, morphologies, and object
sizes that appear in their own data, rather than relying on a single
universal training set. The same applies to the artefact class: each
group can label the artefacts they actually encounter. Both routes are
ultimately bounded by labelling quality: with a $0.49\%$
intra-operator coefficient of variation on the benchmark image,
multi-operator ground truth would be needed to push performance
meaningfully further, but YOLOv8 has proven the capability to locate
these dislocations in ECCI of GaN with high accuracy and speed.

\section{Conclusion}
We benchmarked three automated dislocation detection pipelines on GaN
ECCI micrographs against a statistical ground truth of $3805 \pm 46$
dislocations (95\% CI, $n = 3$). A rule-based pipeline proved
unworkable on dense or strongly contrast-varying images. A multi-stage
CNN reached $87.0 \pm 1.1\%$ counting accuracy but required $\approx
620$~minutes and per-image tuning. A unified YOLOv8 detector with
adaptive tile sizing and a recalibrated confidence threshold reached
$98.6 \pm 1.2\%$ in $\approx 28$\,s for the largest benchmark image
(typically 6--18\,s for smaller images in the test set) with no manual
parameters; aggregate evaluation across 7451 dislocations gave
precision $98.7\%$ and recall $98.7\%$, placing automated counting
within the same precision band as the manual measurement itself.

The residual error modes, faint dipoles near the limits of the chosen
channelling conditions and overlapping dislocations in dense clusters,
are limits of ECCI itself as much as of the detector, and the same
ambiguity propagates into any labels used for training. Pushing
accuracy meaningfully beyond this point likely requires complementary
techniques such as electron backscatter diffraction (EBSD), which can
provide further structural data such as grain orientation and strain
across the sample giving more insight to the crystal but at higher
acquisition cost per image. For routine high-throughput dislocation
density measurement, however, the YOLO-based method described here is
accurate enough to remove manual counting as a bottleneck without
sacrificing the precision of the underlying measurement. A final
caveat concerns verification bias in semi-automated workflows: once
the detector draws a bounding box on a marginal feature, the annotator
is primed to perceive a dislocation that they may not have identified
unaided, and cannot subsequently ``unsee'' it. Manual verification of
detector output is therefore not fully independent of the model under
test, and this anchoring effect should be borne in mind whenever
detector outputs are used to correct counts or to refine training
labels.

\begin{acknowledgments}
We would like to acknowledge Prof.\ Peter Parbrook (University College
Cork and Tyndall National Institute, Ireland), Dr.\ Menno Kappers, and
Prof.\ Rachel Oliver (University of Cambridge, UK) for providing the
GaN samples used in this study. J.B.\ would also like to thank
Dr.\ Grzegorz Cios (AGH University of Krakow, Poland) for providing
access to the SEM facility.

J.B.\ would like to thank the Royal Society of Edinburgh (RSE) for a
Saltire International Collaboration Award [grant number:
  1917]. Prof.\ Daniel Oi is acknowledged for the provision of
computational resources. A.H.\ acknowledges financial support from the
Engineering and Physical Sciences Research Council (EPSRC) [grant
  number: EP/W524670/1].

\end{acknowledgments}

\section*{Data Availability}
The code and trained models supporting this study will be openly
available on GitHub. The data that support the findings of this study
will be available or from the corresponding author upon reasonable
request.

\bibliography{ref}

@book{hull2011introduction,
  title={Introduction to Dislocations},
  author={Hull, D. and Bacon, D. J.},
  edition={5},
  year={2011},
  publisher={Butterworth-Heinemann}
}

@article{Bennett01092010,
author = {S. E. Bennett},
title = {Dislocations and their reduction in {GaN}},
journal = {Materials Science and Technology},
volume = {26},
number = {9},
pages = {1017--1028},
year = {2010},
publisher = {Taylor \& Francis},
doi = {10.1179/026708310X12668415533685},
URL = {https://doi.org/10.1179/026708310X12668415533685}
}

@article{usami2018correlation,
  title={Correlation between dislocations and leakage current of pn diodes on a free-standing {GaN} substrate},
  author={Usami, Shigeyoshi and Ando, Yuto and Tanaka, Atsushi and Nagamatsu, Kentaro and Deki, Manato and Kushimoto, Maki and Nitta, Shugo and Honda, Yoshio and Amano, Hiroshi and Sugawara, Yoshihiro and others},
  journal={Applied Physics Letters},
  volume={112},
  number={18},
  pages={182106},
  year={2018},
  publisher={AIP Publishing},
  doi = {10.1063/1.5024704}
}

@book{meneghini2017power,
  title={Power {GaN} Devices},
  author={Meneghini, Matteo and Meneghesso, Gaudenzio and Zanoni, Enrico},
  year={2017},
  publisher={Springer},
  doi={10.1007/978-3-319-43199-4}
}

@article{SIMKIN199965,
title = {An experimentally convenient configuration for electron channeling contrast imaging},
journal = {Ultramicroscopy},
volume = {77},
number = {1},
pages = {65--75},
year = {1999},
issn = {0304-3991},
doi = {10.1016/S0304-3991(99)00009-1},
url = {https://www.sciencedirect.com/science/article/pii/S0304399199000091},
author = {B. A. Simkin and M. A. Crimp}
}

@article{AKASAKI1989209,
title = {Effects of {AlN} buffer layer on crystallographic structure and on electrical and optical properties of {GaN} and {Ga$_{1-x}$Al$_x$N} ($0 < x \leq 0.4$) films grown on sapphire substrate by {MOVPE}},
journal = {Journal of Crystal Growth},
volume = {98},
number = {1},
pages = {209--219},
year = {1989},
issn = {0022-0248},
doi = {10.1016/0022-0248(89)90200-5},
url = {https://www.sciencedirect.com/science/article/pii/0022024889902005},
author = {Isamu Akasaki and Hiroshi Amano and Yasuo Koide and Kazumasa Hiramatsu and Nobuhiko Sawaki}
}

@article{speck1999mechanisms,
  title={The mechanisms for threading dislocation generation in {GaN}},
  author={Speck, JS and Rosner, SJ},
  journal={Physica B: Condensed Matter},
  volume={273},
  pages={24--32},
  year={1999},
  doi={10.1016/S0921-4526(99)00399-3}
}

@article{nakamura2015nobel,
  title = {{Nobel Lecture}: Background story of the invention of efficient blue {InGaN} light emitting diodes},
  author = {Nakamura, Shuji},
  journal = {Rev. Mod. Phys.},
  volume = {87},
  issue = {4},
  pages = {1139--1151},
  numpages = {13},
  year = {2015},
  month = {Oct},
  publisher = {American Physical Society},
  doi = {10.1103/RevModPhys.87.1139},
  url = {https://link.aps.org/doi/10.1103/RevModPhys.87.1139}
}

@article{Amano_2018,
doi = {10.1088/1361-6463/aaaf9d},
url = {https://doi.org/10.1088/1361-6463/aaaf9d},
year = {2018},
month = {mar},
publisher = {IOP Publishing},
volume = {51},
number = {16},
pages = {163001},
author = {Amano, H and Baines, Y and Beam, E and Borga, Matteo and Bouchet, T and Chalker, Paul R and Charles, M and Chen, Kevin J and Chowdhury, Nadim and Chu, Rongming and De Santi, Carlo and De Souza, Maria Merlyne and Decoutere, Stefaan and Di Cioccio, L and Eckardt, Bernd and Egawa, Takashi and Fay, P and Freedsman, Joseph J and Guido, L and H{\"a}berlen, Oliver and Haynes, Geoff and Heckel, Thomas and Hemakumara, Dilini and Houston, Peter and Hu, Jie and Hua, Mengyuan and Huang, Qingyun and Huang, Alex and Jiang, Sheng and Kawai, H and Kinzer, Dan and Kuball, Martin and Kumar, Ashwani and Lee, Kean Boon and Li, Xu and Marcon, Denis and M{\"a}rz, Martin and McCarthy, R and Meneghesso, Gaudenzio and Meneghini, Matteo and Morvan, E and Nakajima, A and Narayanan, E M S and Oliver, Stephen and Palacios, Tom{\'a}s and Piedra, Daniel and Plissonnier, M and Reddy, R and Sun, Min and Thayne, Iain and Torres, A and Trivellin, Nicola and Unni, V and Uren, Michael J and Van Hove, Marleen and Wallis, David J and Wang, J and Xie, J and Yagi, S and Yang, Shu and Youtsey, C and Yu, Ruiyang and Zanoni, Enrico and Zeltner, Stefan and Zhang, Yuhao},
title = {The 2018 {GaN} power electronics roadmap},
journal = {Journal of Physics D: Applied Physics}
}

@article{fujito2009bulk,
  title={Bulk {GaN} crystals grown by {HVPE}},
  author={Fujito, Kenji and Kubo, Shuichi and Nagaoka, H and Mochizuki, T and Namita, H and Nagao, S},
  journal={Journal of Crystal Growth},
  volume={311},
  number={10},
  pages={3011--3014},
  year={2009},
  doi={10.1016/j.jcrysgro.2009.01.046}
}

@article{mathis2001modeling,
  title={Modeling of threading dislocation reduction in growing {GaN} layers},
  author={Mathis, SK and Hallin, AEN and Speck, JS},
  journal={Journal of Crystal Growth},
  volume={231},
  number={3},
  pages={371--381},
  year={2001},
  doi={10.1016/S0022-0248(01)01468-3}
}

@book{szeliski2010computer,
  title={Computer Vision: Algorithms and Applications},
  author={Szeliski, Richard},
  year={2010},
  publisher={Springer},
  doi={10.1007/978-1-84882-935-0}
}

@article{otsu1979threshold,
  title={A threshold selection method from gray-level histograms},
  author={Otsu, Nobuyuki},
  journal={IEEE Transactions on Systems, Man, and Cybernetics},
  volume={9},
  number={1},
  pages={62--66},
  year={1979},
  doi={10.1109/TSMC.1979.4310076}
}

@article{comaniciu2002mean,
  title={Mean shift: A robust approach toward feature space analysis},
  author={Comaniciu, Dorin and Meer, Peter},
  journal={IEEE Transactions on Pattern Analysis and Machine Intelligence},
  volume={24},
  number={5},
  pages={603--619},
  year={2002},
  doi={10.1109/34.1000236}
}

@article{fukunaga1975estimation,
  title={The estimation of the gradient of a density function, with applications in pattern recognition},
  author={Fukunaga, Keinosuke and Hostetler, Larry},
  journal={IEEE Transactions on Information Theory},
  volume={21},
  number={1},
  pages={32--40},
  year={1975},
  doi={10.1109/TIT.1975.1055330}
}

@article{lecun2015deep,
  title={Deep learning},
  author={LeCun, Yann and Bengio, Yoshua and Hinton, Geoffrey},
  journal={Nature},
  volume={521},
  number={7553},
  pages={436--444},
  year={2015},
  doi={10.1038/nature14539}
}

@inproceedings{redmon2016yolo,
  title={You only look once: Unified, real-time object detection},
  author={Redmon, Joseph and Divvala, Santosh and Girshick, Ross and Farhadi, Ali},
  booktitle={Proceedings of the IEEE Conference on Computer Vision and Pattern Recognition},
  pages={779--788},
  year={2016},
  doi={10.1109/CVPR.2016.91}
}

@inproceedings{wang2020cspnet,
  title={{CSPNet}: A new backbone that can enhance learning capability of {CNN}},
  author={Wang, Chien-Yao and Liao, Hong-Yuan Mark and Wu, Yueh-Hua and Chen, Ping-Yang and Hsieh, Jun-Wei and Yeh, I-Hau},
  booktitle={Proceedings of the IEEE/CVF Conference on Computer Vision and Pattern Recognition Workshops},
  pages={390--391},
  year={2020},
  doi={10.1109/CVPRW50498.2020.00203}
}

@inproceedings{liu2018path,
  title={Path aggregation network for instance segmentation},
  author={Liu, Shu and Qi, Lu and Qin, Haifang and Shi, Jianping and Jia, Jiaya},
  booktitle={Proceedings of the IEEE Conference on Computer Vision and Pattern Recognition},
  pages={8759--8768},
  year={2018},
  doi={10.1109/CVPR.2018.00913}
}

@misc{yolov8_ultralytics,
  title={{YOLO} by {Ultralytics}},
  author={Jocher, Glenn and Chaurasia, Ayush and Qiu, Jing},
  year={2023},
  howpublished={\url{https://github.com/ultralytics/ultralytics}},
  note={Accessed: 2024-01-01}
}

@inproceedings{bendale2016openmax,
  title={Towards open set deep networks},
  author={Bendale, Abhijit and Boult, Terrance E},
  booktitle={Proceedings of the IEEE Conference on Computer Vision and Pattern Recognition},
  pages={1563--1572},
  year={2016},
  doi={10.1109/CVPR.2016.173}
}

@article{student1908,
  title={The probable error of a mean},
  author={Student},
  journal={Biometrika},
  volume={6},
  number={1},
  pages={1--25},
  year={1908},
  doi = {10.2307/2331554}
}

@inproceedings{neubeck2006nms,
  title={Efficient non-maximum suppression},
  author={Neubeck, Alexander and Van Gool, Luc},
  booktitle={18th International Conference on Pattern Recognition (ICPR'06)},
  volume={3},
  pages={850--855},
  year={2006},
  organization={IEEE},
  doi={10.1109/ICPR.2006.479}
}

@article{jaccard1912,
  title={The distribution of the flora in the alpine zone},
  author={Jaccard, Paul},
  journal={New Phytologist},
  volume={11},
  number={2},
  pages={37--50},
  year={1912},
  doi={10.1111/j.1469-8137.1912.tb05611.x}
}

@ARTICLE{mishra2008gan,
  author={Mishra, Umesh K. and Shen, Likun and Kazior, Thomas E. and Wu, Yi-Feng},
  journal={Proceedings of the IEEE}, 
  title={{GaN}-based {RF} power devices and amplifiers}, 
  year={2008},
  volume={96},
  number={2},
  pages={287--305},
  doi={10.1109/JPROC.2007.911060}}

@article{patrick2025comparative,
    author = {Patrick, Matthew J. and Field, Christopher R. and Grae, Lauren H. L. and Rickman, Jeffrey M. and Field, Kevin G. and Barmak, Katayun},
    title = {A comparative analysis of {YOLOv8} and {U-Net} image segmentation approaches for transmission electron micrographs of polycrystalline thin films},
    journal = {APL Machine Learning},
    volume = {3},
    number = {3},
    pages = {036105},
    year = {2025},
    month = {07},
    issn = {2770-9019},
    doi = {10.1063/5.0274266},
    url = {https://doi.org/10.1063/5.0274266}
    }

@Article{Rafin2023,
AUTHOR = {Rafin, S M Sajjad Hossain and Ahmed, Roni and Haque, Md. Asadul and Hossain, Md. Kamal and Haque, Md. Asikul and Mohammed, Osama A.},
TITLE = {Power electronics revolutionized: A comprehensive analysis of emerging wide and ultrawide bandgap devices},
JOURNAL = {Micromachines},
VOLUME = {14},
NUMBER = {11},
PAGES = {2045},
YEAR = {2023},
ISSN = {2072-666X},
DOI = {10.3390/mi14112045}
}

@ARTICLE{Paskova2010,
  author={Paskova, Tanya and Hanser, Drew A. and Evans, Keith R.},
  journal={Proceedings of the IEEE}, 
  title={{GaN} substrates for {III}-nitride devices}, 
  year={2010},
  volume={98},
  number={7},
  pages={1324--1338},
  doi={10.1109/JPROC.2009.2030699}}

@InProceedings{Lin_2017_CVPR,
author = {Lin, Tsung-Yi and Dollar, Piotr and Girshick, Ross and He, Kaiming and Hariharan, Bharath and Belongie, Serge},
title = {Feature pyramid networks for object detection},
booktitle = {Proceedings of the IEEE Conference on Computer Vision and Pattern Recognition (CVPR)},
month = {July},
year = {2017},
doi = {10.1109/CVPR.2017.106}
}

@article{moram2009origin,
    author = {Moram, M. A. and Ghedia, C. S. and Rao, D. V. S. and Barnard, J. S. and Zhang, Y. and Kappers, M. J. and Humphreys, C. J.},
    title = {On the origin of threading dislocations in {GaN} films},
    journal = {Journal of Applied Physics},
    volume = {106},
    number = {7},
    pages = {073513},
    year = {2009},
    month = {10},
    issn = {0021-8979},
    doi = {10.1063/1.3225920},
    url = {https://doi.org/10.1063/1.3225920}
}

@article{joy1982electron,
    author = {Joy, David C. and Newbury, Dale E. and Davidson, David L.},
    title = {Electron channeling patterns in the scanning electron microscope},
    journal = {Journal of Applied Physics},
    volume = {53},
    number = {8},
    pages = {R81--R122},
    year = {1982},
    month = {08},
    issn = {0021-8979},
    doi = {10.1063/1.331668},
    url = {https://doi.org/10.1063/1.331668}
}

@article{wilkinson1997electron,
title = {Electron diffraction based techniques in scanning electron microscopy of bulk materials},
journal = {Micron},
volume = {28},
number = {4},
pages = {279--308},
year = {1997},
issn = {0968-4328},
doi = {10.1016/S0968-4328(97)00032-2},
url = {https://www.sciencedirect.com/science/article/pii/S0968432897000322},
author = {Angus J. Wilkinson and Peter B. Hirsch}
}

@Article{hiller2026imaging,
author = {Hiller, Kieran P. and Cios, Grzegorz and Winkelmann, Aimo and Wheeler, John and Parbrook, Peter J. and Hourahine, Ben and Trager-Cowan, Carol and Bruckbauer, Jochen},
journal = {Acta Materialia},
title = {Imaging misorientation and strain of single dislocations in {GaN} using electron backscatter diffraction},
year = {2026},
issn = {1359-6454},
month = {April},
pages = {122185},
volume = {312},
doi = {10.1016/j.actamat.2026.122185},
url = {https://strathprints.strath.ac.uk/95965/},
}

@article{li2018automated,
  title={Automated defect analysis in electron microscopic images},
  author={Li, Wei and Field, Kevin G and Morgan, Dane},
  journal={npj Computational Materials},
  volume={4},
  pages={36},
  year={2018},
  doi={10.1038/s41524-018-0093-8}
}

@article{roberts2019defectsegnet,
  title={Deep learning for semantic segmentation of defects in advanced {STEM} images of steels},
  author={Roberts, Graham and Haile, Simon Y and Sainju, Rajat and Edwards, Danny J and Hutchinson, Brian and Zhu, Yuanyuan},
  journal={Scientific Reports},
  volume={9},
  pages={12744},
  year={2019},
  doi={10.1038/s41598-019-49105-0}
}

@article{shen2021multidefect,
  title={Multi defect detection and analysis of electron microscopy images with deep learning},
  author={Shen, Mingren and Li, Guanzhao and Wu, Dongxia and Liu, Yuhan and Greaves, Jacob R C and Hao, Wei and Krakauer, Nathaniel J and Krudy, Leah and Perez, Jacob and Sreenivasan, Varun and others},
  journal={Computational Materials Science},
  volume={199},
  pages={110576},
  year={2021},
  doi={10.1016/j.commatsci.2021.110576}
}

\end{document}